\documentstyle[12pt]{article}
\makeatletter \oddsidemargin  -.1in \evensidemargin -.1in
\newcommand{\singlespacing}{\let\CS=\@currsize\renewcommand{\baselinestretch}{1}\tiny\CS}
\newcommand{\oneandahalfspacing}{\let\CS=\@currsize\renewcommand{\baselinestretch}{1.25}\tiny\CS}
\newcommand{\doublespacing}{\let\CS=\@currsize\renewcommand{\baselinestretch}{1.35}\tiny\CS}

\newtheorem{rule-def}[theorem]{Rule}

\RequirePackage[dvips]{graphicx} \textheight 22.5cm
\begin{document}         
\title {\bf Peristaltic Flow of a Fluid in a Porous Channel: a Study Having Relevance to Flow of Bile Within Ducts in a Pathological State}
\author{\small ~S. Maiti$^1$\thanks{Email address:
    {\it somnathm@cts.iitkgp.ernet.in (S. Maiti)}}~, J. C. Misra$^2$\thanks{Email address:
    {\it misrajc@rediffmail.com; jcm@maths.iitkgp.ernet.in
      (J.C.Misra)}}~  \\ $^1$\it School of Medical Science and
  Technology $\&$ Centre for Theoretical Studies,\\\it Indian Institute
  of Technology, Kharagpur, India\\
\it$^2$Department of Mathematics, Institute of Technical
Education $\&$ Research,\\\it Siksha 'O' Anusandhan University, Bhubaneswar, India}
\date{}
\maketitle \noindent \doublespacing
\begin{abstract}
The paper deals with a theoretical study of the transport of a fluid
in a channel, which takes place by the phenomenon of peristalsis. A
mathematical analysis of the said problem has been presented. The
analysis involves the application of a suitable perturbation
technique. The velocity profile and the critical pressure for the
occurrence of reflux are investigated with particular emphasis by
using appropriate numerical methods. The effects of various
parameters, such as Reynolds number, pressure gradient, porosity
parameter, Darcy number, slip parameter, amplitude ratio and wave
number on velocity and critical pressure for reflux are investigated
in detail. The computed results are compared with a previous
analytical work and an experimental investigation reported earlier in
existing scientific literatures. The results of the present study are
in conformity to both of them. The study has got some relevance to the
physiological flow of bile in the common bile duct in a pathological
state. It reveals that in the presence of gallstones, bile velocity
increases as the value of the porosity parameter increases, while the
critical pressure for reflux decreases as porosity increases.
\end{abstract}

\section{Introduction}
 Peristaltic pumping \cite{r1, r2} of physiological fluids induced by a
progressive wave of area contraction or expansion along the length of
a distensible tube has drawn serious attention of researchers working
in the area of physiological fluid dynamics. Movement of various
physiological fluids, such as the transport of urine from the kidney
to the bladder through ureters, movement of food material through the
digestive tract, transport
\begin{center}
\begin{tabular}{|l l|}\hline
{~\bf Nomenclature} &~ \\
~~$a$ & Traveling wave amplitude\\
~~$c_{20},c_{21},c_{22} $ & Integration constants\\
~~$d $ & Half width of mean spacing between solid boundaries \\
~~$D $ & Constant defined in equation (34)\\
~~$f(y) $ & Function defined in equation (35)\\
~~$k $ & Permeability parameter/Darcy number \\
~~$p $ & Fluid pressure\\
~~$\frac{\partial p}{\partial x} $ & Pressure gradient averaged over a period of time\\
~~$R $ & Reynolds number of the fluid\\
~~$s $ & Slip parameter at the boundary\\
~~$t $  & Time\\
~~$u,v$ & Velocity components in X and Y directions respectively\\
~~$x,y $ & Rectangular Cartesian co-ordinates\\
~~$\alpha $ & Wave number\\
~~$\beta $ & Complex number defined in equation (31)\\
~~$e $ & Porosity parameter \\
~~$\lambda $ & Wave length of the travelling wave motion of the wall\\
~~$\mu $ & Dynamic viscosity of the fluid\\
~~$\mu_1 $ & Apparent viscosity of the fluid\\
~~$\phi$ & Phase difference\\
~~$\rho$ & Density of the fluid \\
~~$\phi_1,\phi_{20},\phi_{22} $ & Functions defining $\psi_1,\psi_2$ in equation (20)and (21) respectively\\
~~$\psi$ & Stream function\\
~~$\epsilon $ & Amplitude ratio\\
~~$\eta $ & Vertical displacements of the wall\\
\hline
\end{tabular}
\end{center}
of semen in the vas differences, fluids movement of lymphatic fluids
in lymphatic vessels, flow of bile (cf. Daniel et al. \cite{r3}) from
gall bladder into the duodenum, spermatozoa in the ductus efferentes
of the male reproductive tract and cervical canal, movement of ovum in
the fallopian tube as well as the cilia movement and circulation of
blood in small blood vessels, --- are all performed by the mechanism
of peristalsis. The propulsion of some industrial fluids are take
place by this mechanism. Peristaltic transport has also found various
applications in roller and finger pumps, heart-lung machines, blood
pump machines, dialysis machines and also transport of noxious fluid
in nuclear industries.

 A few theoretical studies on peristaltic transport of different
 physiological fluids were carried out by Usha and Rao \cite{r4},
 Mishra and Rao \cite{r5,r6}, Misra et al
 \cite{r7,r8,r9,r10,r11,r12}, as well as Eytan et al. \cite{r13}. The
 analyses were mostly restricted to consideration of small peristaltic
 wave amplitude and the assumption that the fluid inertia is
 negligible.

Taylor \cite{r14} conducted an investigation on asymmetric wave
propagation in wavy sheets with the main objective of deriving some
information regarding the mechanical interaction between
spermatozoa. In order to study some fluid dynamical aspect of the
problem of peristaltic transport in an asymmetric channel under Stokes
flow conditions, the boundary integral method along with a suitable
numerical technique was employed by Pozrikidis \cite{r15}. Using the
assumptions of thin shell and lubrication theories, Carew and Pedley
\cite{r16} put forward a mathematical analysis for the development of
peristaltic pumping in the ureter by using similar assumptions. By
taking into account the wall deformation of the pipe, Antanovskii and
Ramkissoon \cite{r17} also used lubrication theory in order to analyse
the peristaltic motion of a compressible viscous fluid through a pipe
for situations where the pressure drop changes with time.

Bile (alternatively called as gall) is a greenish yellow secretion that
contains various biochemical substances like bile acids, bile salts,
pigments, cholesterol, phospholipids and electrolytic chemicals. Bile
is produced in the liver. In adult humans, the quantity produced in a
day is about one litre. After passing through several bile ducts,
which penetrate the liver, bile flows into to the the common bile
duct. It helps accelerate the fat absorption process and there by
plays an important role in absorbing the vitamins D, E, K and A that
are soluble in fat.

It has been suggested by recent physiological researchers (cf. Vries
et al.\cite{r18} ) that uterine peristalsis resulting from myometrial
contraction can take place in both symmetric and asymmetric
directions. Various investigators have carried out different studies
pertaining to the gastrointestinal tract, intra-pleural membranes,
capillary walls, human lung, bile duct, gall bladder with stones and
small blood vessels, as well as flow in porous tubes and deformable
porous layers. Keener and Sneyd \cite{r19} reported that
gastrointestinal tract is surrounded by a number of heavily innervated
muscle layers which are smooth muscles consisting of many folds. There
exist pores in the junctions between them, although the junctions are
tight. As mentioned by Bergel \cite{r20}, the capillary walls are
surrounded by flattened endothelial cell layers, which are porous. Li
et al. \cite{r21} reported that an impulsive magnetic field can be
used as a theraptic means to treat patients who have stone fragments
in their urinary tract.

Functions of the human biliary system that consists of an organ and a
ductal system are to create, transport, store as well as release bile
into the duodenum to assist digestion of fats. This system contains
the liver, gallbladder and biliary tract namely cystic, hepatic and
common bile ducts. Although several analytical and physiological
aspects of the human biliary system have been analyzed elaborately, we
have only insufficient information about the mechanism of bile flow in
the system. Torosoli and Ramorino \cite{r22} reported that pressures
in the biliary tree vary from 0-14 cm $H_2O$ (1 cm $H_2O=100$ Pa) in
the resting gall bladder to approximately 12-20 cm $H_2O$ in the
common bile duct.
 
Cholelithiasis is a disease that concerns formation of Gallstones. It
has become a major health problem worldwide, particularly in adult
population. Incidence of the gallstone disease indicates considerable
geographical and regional variations \cite{r23,r24,r25,r26,r27,r28}.

Lauga and Stone \cite {r29} made an experimental attempt to evaluate
the effective slip length of the resulting flow as a function of the
degrees of freedom describing the surface heterogeneities, namely the
relative width of the no-slip and no-shear stress regions and their
distribution along the pipe. They gave a possible interpretation of
the experimental results which is consistent with a large number of
distributed slip domains such as nano-size and micron-size nearly flat
bubbles coating the solid surface. In addition they also suggested
that the possibility of a shear-dependent effective slip length.

There are, however, only a few studies on the rheological properties
of human bile. Gottschack and Lochner \cite {r30} examined 33 samples
and reported that post-operative T-tube human bile is a Maxwell
fluid. Coene et al. \cite {r31} claimed that 11 of 36 hepatic bile
samples displayed non-Newtonian behaviour. Lou et al. \cite {r32}
suggested on the basis of their preliminary measurements of fresh
human bile collected at $37^\circ$C for a healthy person without
gallstones that bile is a Newtonian fluid.
                 
From the physiological fluid dynamics point of view, we present here a
theoretical study that has relevance to the problem of bile transport
in the common bile duct in the presence of stones. Peristaltic
transport of the fluid in a porous channel has been investigated. The
symmetric sinusoidal peristaltic wave train on the channel walls,
impermeable boundary and slip boundary condition of Saffman type have
been considered in this study. The overall aim of the study has been
to examine the role of fluid dynamics in the human biliary system, in
particular for flow in a common bile duct with/without stones. It is
expected that the results presented here will serve as fairly good
theoretical estimates of various prospective fluid mechanical flow
governing parameters related to the peristaltic transport of bile.

\section{Problem Formulation}

\begin{minipage}{1.0\textwidth}
     \begin{center}
       \includegraphics[width=3.25in,height=2.0in]{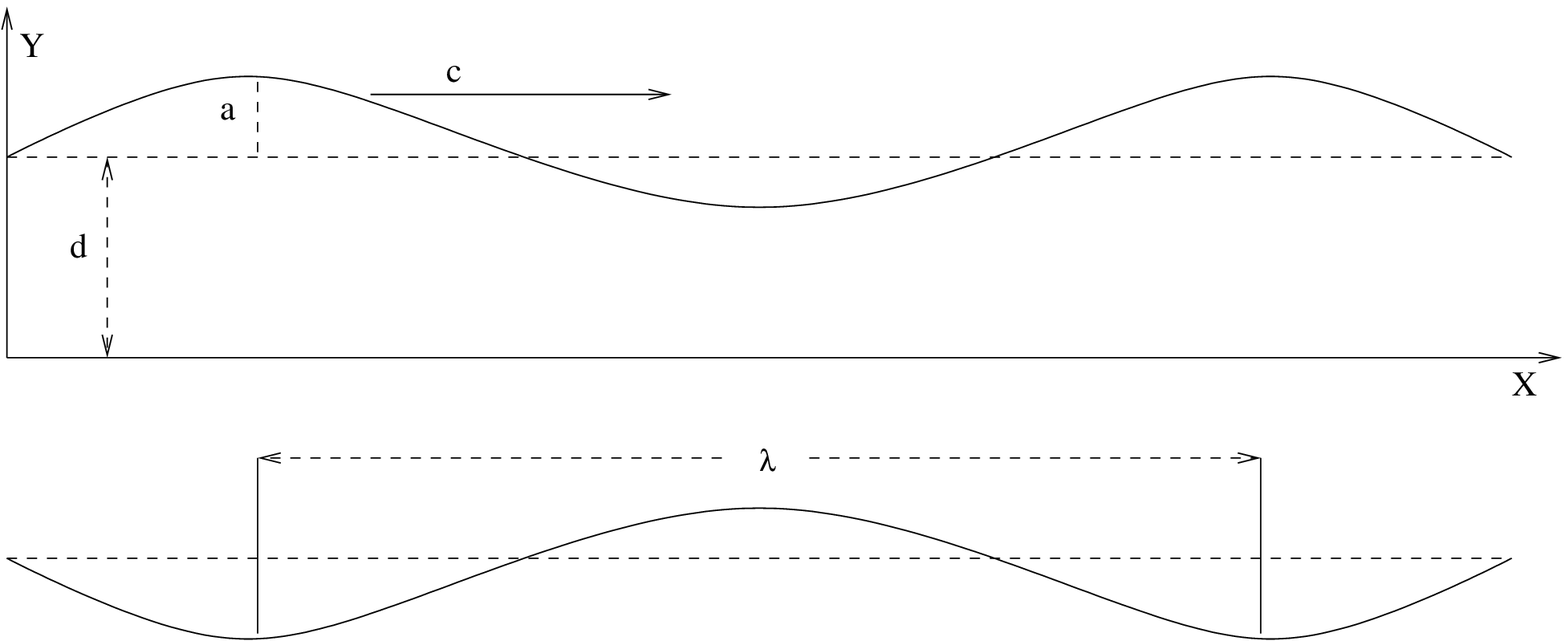} \\
 Fig. 1: Geometry of the problem \\
\end{center}
\end{minipage}\vspace*{.25cm}
In a pathological state, numerous stones are formed in the bile
(physiological fluid). In such a situation the mixture of the fluid
(bile) and the solid particles (stones) forms a dense porous mass. We
are interested here to study the motion of such a fluid (dense
mixture), by considering it as a porous mass and treating it as an
incompressible viscous fluid. Slip boundary conditions of Saffman's
type have been considered in the study. We use Cartesian coordinates
(x,y), where x is measured in the direction of wave propagation and y
in the direction normal to the mean position of the corresponding
common bile duct (cf. Fig.1). Let u, v denote the velocity components
in the directions of increasing x and y respectively; $\rho$, p, $\nu$
and t stand for the density, pressure, kinematic viscosity of the
fluid and time respectively. A sinusoidal wave train is propagating
with a constant speed $c$ along the channel wall, such that $\eta=a
cos(\frac{2\pi}{\lambda}(x-ct)) $where $ y=d+\eta$ and $y=-d-\eta$
represent the upper and lower boundaries of the channel and $a$ is the
wave amplitude, $\lambda$ the wave length, 2d the width of the
channel.

In order to study the motion of the fluid in the porous structure, we
make use of Brinkman's equations. These equations and the equation of
continuity may be put as
\begin{equation}
\rho\left(\frac{\partial u}{\partial t}+u\frac{\partial u}{\partial x}+v\frac{\partial u}{\partial y}\right)=-\frac{\partial p}{\partial x}+\mu_1 \left (\frac{\partial^2 u}{\partial x^2}+\frac{\partial^2 u}{\partial y^2}\right )-\frac{\mu}{k}u
\end{equation}
\begin{equation}
\rho\left(\frac{\partial v}{\partial t}+u\frac{\partial v}{\partial
x}+v\frac{\partial v}{\partial y}\right)=-\frac{\partial
p}{\partial y}+\mu_1 \left (\frac{\partial^2 v}{\partial
x^2}+\frac{\partial^2 v}{\partial y^2}\right )-\frac{\mu}{k}v
\end{equation}
\begin{equation}
\frac{\partial u}{\partial x}+\frac{\partial v}{\partial y}=0
\end{equation}
where $\mu_1=\frac{\mu}{e}$ and e, k are respectively the porosity, permeability parameters. \\

Introducing the stream function $\psi$, we write\\ $u=\frac{\partial \psi}{
\partial y}, ~~ v=-\frac{\partial \psi}{\partial x} $.\\
 Eliminating p, one can write the equation that governs the flow of the fluid in terms of $\psi$ in the following form : 
 \begin{eqnarray}
 \psi_{tyy}+\psi_{txx}+\psi_y\psi_{yyx}-\psi_x\psi_{yyy}+ \psi_y\psi_{xxx}-
\psi_x\psi_{xxy} &=& \frac{\nu}{e}\left(\psi_{yyyy}+2\psi_{xxyy}+\psi_{xxxx}\right) \nonumber \\  &-&\frac{\nu}{k}\psi_{yy}-\frac{\nu}{k}\psi_{xx},~in~which~\nu=\frac{\mu}{\rho} 
\end{eqnarray}
In the sequel, we shall use the following non-dimensional variables
defined by
\begin{eqnarray}
&\bar{x}&=\frac{x}{d}, ~~\bar{y}=\frac{y}{d}, ~~\bar{u}=\frac{u}{c},
~~\bar{v}=\frac{v}{c}, ~~ \epsilon=\frac{a}{d}, ~~\bar{p}=\frac{p}{\rho c^2},~~
\bar{t}=\frac{ct}{d}, ~~\bar{\eta}=\frac{\eta}{d},\nonumber \\
~~& R&=\frac{c d}{\nu }, ~~ \bar{\psi}=\frac{\psi}{cd}, ~~\bar{k}=\frac{k}{d^2},~~\alpha=\frac{2 \pi d}{\lambda}
\end{eqnarray}
  If we now drop bars over the symbols, equation (4) may be expressed as
 \begin{eqnarray}
R(\psi_{tyy}+\psi_{txx}+\psi_y\psi_{yyx}-\psi_x\psi_{yyy}+ \psi_y\psi_{xxx}-
\psi_x\psi_{xxy}) &=& \frac{1}{e} (\psi_{yyyy}+2\psi_{xxyy}+\psi_{xxxx}) \nonumber \\  &-&\frac{1}{k}\psi_{yy}-\frac{1}{k}\psi_{xx} 
\end{eqnarray}
\subsection{Boundary Conditions }
  Let us now consider the symmetric motion of the flexible walls. The
  boundary conditions for the present problem may be stated as follows : 
\begin{eqnarray*}
(i)~non~zero~velocity~slip~of~Saffman~type:~~ u=\mp su_y ~~or ~\psi_y=\mp s\psi_{yy}~at~ y=\pm d \pm \eta~~~~~~~~~~~ \nonumber\\
(ii)~symmetric~motion~of~the~wall:~ v= \pm \frac{\partial \eta}{\partial t}~
or-\psi_x=\pm \frac{2\pi ac}{\lambda} \sin\left(\frac{2\pi}{\lambda}(x-ct)\right)~ at~ y=\pm d \pm \eta \nonumber
\end{eqnarray*}
\begin{equation}
\end{equation}

\section{Solution Procedure}
In order to solve the governing equation along with the boundary
conditions, let us express the stream function $\psi$ as a power
series in terms of amplitude ratio $\epsilon$ (considered as a small
quantity), as follows :
\begin{eqnarray}
\psi=\psi_0+\epsilon \psi_1+\epsilon^2 \psi_2+... 
\end{eqnarray} 
Similarly, we write the pressure gradient $\frac{\partial p}{\partial x}$ as
\begin{eqnarray}
\frac{\partial p}{\partial x}=\left(\frac{\partial p}{\partial x}\right)_0+\epsilon \left(\frac{\partial p}{\partial x}\right)_1+\epsilon^2 \left(\frac{\partial p}{\partial x}\right)_2+...
\end{eqnarray}
In equation (9) the first term on the right corresponds to an imposed
pressure gradient, while the higher order terms are conceived of as
those arising out of the peristaltic mechanism. Substituting (8) into
(6), equating the coefficients of like powers of $\epsilon$ and
neglecting cubes and higher power of $\epsilon$, we get
\begin{equation}
\frac{\partial }{\partial t}\nabla^2 \psi_0+\psi_{0y}\nabla^2\psi_{0x}-\psi_{0x}\nabla^2\psi_{0y}=\frac{1}{eR}\nabla^2\nabla^2\psi_0-\frac{1}{kR}\nabla^2\psi_0
\end{equation}
\begin{equation}
\frac{\partial }{\partial t}\nabla^2 \psi_1+\psi_{1y}\nabla^2\psi_{0x}+\psi_{0y}\nabla^2\psi_{1x}-\psi_{1x}\nabla^2\psi_{0y}-\psi_{ox}\nabla^2\psi_{1y}=\frac{1}{eR}\nabla^2\nabla^2\psi_1-\frac{1}{kR}\nabla^2\psi_1
\end{equation}
\begin{equation}
\frac{\partial }{\partial t}\nabla^2 \psi_2+\psi_{2y}\nabla^2\psi_{0x}+\psi_{1y}\nabla^2\psi_{1x}+\psi_{0y}\nabla^2\psi_{2x}-\psi_{2x}\nabla^2\psi_{0y}-\psi_{1x}\nabla^2\psi_{1y}-\psi_{0x}\nabla^2\psi_{2y}=\frac{1}{eR}\nabla^2\nabla^2\psi_2-\frac{1}{kR}\nabla^2\psi_2
\end{equation}
in which $\nabla^2=\frac{\partial^2 }{\partial x^2}+\frac{\partial^2 }{\partial y^2}~and~\psi_{ix}=\frac{\partial \psi_{i}}{\partial x},~\psi_{iy}=\frac{\partial \psi_{i}}{\partial y},~i=0,1,2.$~\\
Substituting of (8) into (7), we obtain the following boundary conditions by equating the coefficients of like power of $\epsilon$.
 \begin{eqnarray}
 \psi_0(\pm 1)=0~~and~~\psi_{0y}(\pm 1)=\mp s\psi_{0yy}(\pm 1)
\end{eqnarray}
\begin{eqnarray}
 \psi_{1y}(\pm 1) \pm\psi_{0yy}(\pm 1)\cos\alpha(x-t) =\mp s \{\psi_{1yy}(\pm 1) \pm\psi_{0yyy}(\pm 1)\cos\alpha(x-t)\}\\
 \psi_{1x}(\pm 1) \pm\psi_{0xy}(\pm 1)\cos\alpha(x-t)=\mp \sin\alpha(x-t)~~~~~~~~~~~~~~~~~~~~~~~~~~~~~~~~  
\end{eqnarray}
\begin{eqnarray}
 \psi_{2y}(\pm 1) \pm\psi_{1yy}(\pm 1)\cos\alpha(x-t)+\frac{1}{2}\cos^2\alpha(x-t)\psi_{0yyy}(\pm 1)\nonumber~~~~~~~~~~~~~~~~~~~\\ =\mp s \{\psi_{2yy}(\pm 1) \pm\psi_{1yyy}(\pm 1)\cos\alpha(x-t)+\frac{1}{2}\cos^2\alpha(x-t)\psi_{0yyy}(\pm 1)\}\\
 \psi_{2x}(\pm 1) \pm\psi_{1xy}(\pm 1)\cos\alpha(x-t)+\frac{1}{2}\cos^2\alpha(x-t)\psi_{0xyy}(\pm 1)=0~~~~~~~~~~~~~
\end{eqnarray}
Considering symmetry and a uniform pressure gradient in the
x-direction, the solution of the first set of differential equations
lead to the following classical Poiseuille flow equation that may be
written as
\begin{equation}
\psi_0=A_0\{-C_0 y+\sinh\frac{\sqrt{e}y}{\sqrt{k}},\}
\end{equation}
\begin{eqnarray}
where~~~~~ C_0=\frac{\sqrt{e}}{\sqrt{k}}\cosh\frac{\sqrt{e}}{\sqrt{k}}+\frac{se}{k}\sinh\frac{\sqrt{e}}{\sqrt{k}},~ A_0=\frac{kR\left(\frac{\partial p}{\partial x}\right)_0}{C_0}.
\end{eqnarray}
Let us take the solutions of the differential equations (11) and (12),
which satisfy the respective boundary conditions (14), (15) and (16),
(17) given by
\begin{eqnarray}
2\psi_1(x,y,t)=\phi_1(y)e^{i\alpha (x-ct)}+\phi_1^{\ast}(y)e^{-i\alpha (x-ct)}\\2\psi_2(x,y,t)=\phi_{20}(y)+\phi_{22}(y)e^{2i\alpha (x-ct)}+\phi_{22}^{\ast}(y)e^{-2i\alpha (x-ct)},
\end{eqnarray}
where the asterisk sign denotes the complex conjugate of the
corresponding quantity. We now substitute (20) and (21) into the
differential equations (11) and (12) and the boundary conditions (14),
(15) and (16), (17). Thus we obtain
\begin{eqnarray}
\left\{\frac{d^2 }{d y^2}-\alpha^2-\frac{e}{k}+i \alpha e R\left(1-A_0\left(-C_0+\frac{\sqrt{e}}{\sqrt{k}}\cosh\frac{\sqrt{e}y}{\sqrt{k}}\right)\right)\right\}\left\{\frac{d^2 }{d y^2}-\alpha^2\right\}\phi_1&&\nonumber\\+\frac{i \alpha Re^\frac{5}{2}A_0}{k\sqrt{k}}\cosh\frac{\sqrt{e}y}{\sqrt{k}}\phi_1=0
\end{eqnarray}
\begin{eqnarray}
with~~~\phi_1^\prime(\pm 1)+\frac{e A_0}{k}\sinh\frac{\sqrt{e}}{\sqrt{k}}=\mp s\phi_1^{\prime\prime}-\frac{se^\frac{3}{2}A_0}{k\sqrt{k}}\cosh\frac{\sqrt{e}}{\sqrt{k}},~~~~~~~~~~~~~~~~~~~~~\\
\phi_1(\pm 1)=\pm 1~~~~~~~~~~~~~~~~~~~~~~~~~~~~~~~~~~~~~~~~~~~~~~~~~~~~~~~~~~~~~~~~~
\end{eqnarray}
\begin{eqnarray}
and~~~~~\phi_{20}^{iv}-\frac{e}{k}\phi_{20}^{\prime\prime}&=&-\frac{i\alpha e R}{2}\left(\phi_1\phi_1^{\ast \prime\prime}- \phi_1^\ast \phi_1^{ \prime\prime}\right)^\prime
\end{eqnarray}
\begin{eqnarray}
\left\{\frac{d^2 }{d y^2}-4\alpha^2-\frac{e}{k}+2i\alpha e R\left(1-A_0\left(-C_0+\frac{\sqrt{e}}{\sqrt{k}}\cosh\frac{\sqrt{e}y}{\sqrt{k}}\right)\right)\right\}\left\{\frac{d^2 }{d y^2}-4\alpha^2\right\}\phi_{22}&&\nonumber\\+\frac{2i\alpha Re^\frac{5}{2}A_0}{k\sqrt{k}}\cosh\frac{\sqrt{e}y}{\sqrt{k}}\phi_{22}=\frac{i\alpha e R}{2}\left(\phi_1^\prime \phi_1^{\prime\prime}- \phi_1 \phi_1^{ \prime\prime\prime}\right)
\end{eqnarray}
along with the conditions
\begin{eqnarray}
\phi_{20}^\prime(\pm 1)\pm\frac{1}{2}\left(\phi_1^{\prime\prime}(\pm 1)+\phi_1^{\ast\prime\prime}(\pm 1)\right)+\frac{A_0e^\frac{3}{2}}{2k\sqrt{k}}\cosh\frac{\sqrt{e}}{\sqrt{k}}\nonumber ~~~~~~~~~~~~~~~~~~~\\=\mp s\left\{\phi_{20}^{\prime\prime}(\pm 1)\pm\frac{1}{2}\left(\phi_1^{\prime\prime\prime}(\pm 1)+\phi_1^{\ast\prime\prime\prime}(\pm 1)\right)+\frac{A_0e^2}{2k^2}\sinh\frac{\sqrt{e}}{\sqrt{k}}\right\}\\
\phi_{22}^\prime(\pm 1)\pm\frac{1}{2}\left(\phi_1^{\prime\prime}(\pm 1)+\phi_1^{\ast\prime\prime}(\pm 1)\right)+\frac{A_0e^\frac{3}{2}}{4k\sqrt{k}}\cosh\frac{\sqrt{e}}{\sqrt{k}}\nonumber ~~~~~~~~~~~~~~~~~~~~~\\=\mp s\left\{\phi_{22}^{\prime\prime}(\pm 1)\pm\frac{1}{2}\left(\phi_1^{\prime\prime\prime}(\pm 1)+\phi_1^{\ast\prime\prime\prime}(\pm 1)\right)+\frac{A_0e^2}{4k^2}\sinh\frac{\sqrt{e}}{\sqrt{k}}\right\}\\
\phi_{22}(\pm 1)\pm\frac{1}{4}\phi_1^\prime(\pm 1)=0~~~~~~~~~~~~~~~~~~~~~~~~~~~~~~~~~~~~~~~~~~~~~~~~~~~~~~~~~
\end{eqnarray}
In the above-written equations, primes denote derivatives of the function with respect to 'y' and R denotes Reynolds number.
 
By solving the set of differential equations presented above
along with the respective boundary conditions, it is possible to
obtain the solution of the problem (up to the second order in
$\epsilon$). In the general case, it is difficult to solve the
fourth-order differential equations analytically. It is,
however, possible to find a closed form solution of the equation for
the particular case of pumping of an initially stagnant fluid when
there is no imposed pressure gradient, that is $\left(\frac{\partial
  p}{\partial x}\right)_0=0.$ In this case, $A_0=0$ and the other
coefficients are constants. Such a situation occurs in practice, when
the fluid is stationary and there is no travelling peristaltic
wave. Moreover, the maximum pressure gradient that can be created by a
small amplitude is of order $\epsilon^2$. In the pumping range , the
zeroth order mean pressure gradient must disappear. Therefore, the
consideration is not much restrictive.

Considering $A_0=0$, the solution of equation (22), subject to the
boundary conditions (23) and (24) is found as
\begin{equation}
\phi_1(y) =C_{11}\sinh \alpha y+C_{12}\sinh \beta y,
\end{equation}
in which
\[C_{11}=-\frac{\beta \cosh \beta +s\beta^2 \sinh \beta}{\alpha \cosh\alpha \sinh \beta -\beta \sinh \alpha \cosh \beta +s(\alpha^2-\beta^2)\sinh \alpha \sinh \beta} \]
\[C_{12}=\frac{\alpha \cosh \alpha +s\alpha^2 \sinh \alpha}{\alpha \cosh\alpha \sinh \beta -\beta \sinh \alpha \cosh \beta +s(\alpha^2-\beta^2)\sinh \alpha \sinh \beta} \]
\begin{eqnarray} 
where~~~\beta^2=\alpha^2 +\frac{e}{k}-2i\alpha e R,
\end{eqnarray} 
when  $-d-\eta \le y \le d+\eta .$ 

In order to determine the mean flow, we would need only the term $\phi_{20}^\prime$ in the expansion of $\psi_2$ given by (21). Substituting (30) into (25) and integrating w.r.to 'y' we obtain 
\begin{eqnarray}
\phi_{20}^{\prime\prime\prime}-\frac{e}{k}\phi_{20}^{\prime}=-\frac{i\alpha R}{2}\left [ \left (i \alpha e R -\frac{e}{k}\right)C_{11}^\ast C_{12}\sinh \alpha y \sinh \beta y+\left (i \alpha e R +\frac{e}{k}\right)C_{11} C_{12}^\ast \sinh \alpha y \sinh \beta^\ast y\right.\nonumber\\\left.+2 i \alpha e R C_{12} C_{12}^\ast \sinh \beta y \sinh \beta^\ast y\right ]+2 C_{20},~~~~~~~~~~~~~~~~~~~~~~~~~~~~~~~~~~~~~~~~~~~~~~~~~~~~~~~~
\end{eqnarray}
where $C_{20}$ is an arbitrary constant. The equation (32) further leads to 
\begin{equation}
\phi_{20}^{\prime} =f(y)+\frac{\left \{D-f(1)-sf^{\prime} (1) \right\} \cosh \frac{\sqrt{e}y}{\sqrt{k}}}{ \cosh \frac{\sqrt{e}}{\sqrt{k}}+\frac{s\sqrt{e}}{\sqrt{k}} \sinh \frac{\sqrt{e}}{\sqrt{k}}}-\frac{2C_{20}k}{e} \left ( 1-\frac{\cosh \frac{\sqrt{e}y}{\sqrt{k}}}{ \cosh \frac{\sqrt{e}}{\sqrt{k}}+\frac{s\sqrt{e}}{\sqrt{k}}\sinh \frac{\sqrt{e}}{\sqrt{k}}} \right )
\end{equation}
in which
\begin{eqnarray}
D = \phi_{20}^\prime(\pm 1) \pm s\phi_{20}^{\prime \prime}(\pm 1)
&=&-\frac{1}{2}\left [  \alpha^2 (C_{11}+C_{11}^\ast)\sinh \alpha +\beta^2 C_{12} \sinh \beta +\beta^{\ast 2}  C_{12}^\ast \sinh \beta^{\ast }\right. \nonumber\\ &&\left.+s\left \{ \alpha^3 (C_{11}+C_{11}^\ast)\cosh \alpha +\beta^3 C_{12} \cosh \beta +\beta^{\ast 3} C_{12}^\ast \sinh \beta^{\ast }  \right \} \right ]~~~~~
\end{eqnarray}
\begin{eqnarray}
and~f(y) = -\frac{i \alpha e R}{4}\left [ \left (i \alpha e R -\frac{e}{k}\right)C_{11}^\ast C_{12} \left \{ \frac{\cosh (\alpha +\beta)y}{(\alpha +\beta)^2-\frac{e}{k}}-\frac{\cosh (\alpha -\beta)y}{(\alpha -\beta)^2-\frac{e}{k}}\right \}\right. \nonumber\\  +\left (i \alpha e R +\frac{e}{k}\right)C_{11} C_{12}^\ast \left \{ \frac{\cosh (\alpha +\beta^\ast)y}{(\alpha +\beta^\ast)^2-\frac{e}{k}}-\frac{\cosh (\alpha -\beta^\ast)y}{(\alpha -\beta^\ast)^2-\frac{e}{k}}\right \}  \nonumber\\ \left.  + 2 i \alpha e R C_{12} C_{12}^\ast \left \{ \frac{\cosh (\beta +\beta^\ast)y}{(\beta +\beta^\ast)^2-\frac{e}{k}}-\frac{\cosh (\beta -\beta^\ast)y}{(\beta -\beta^\ast)^2-\frac{e}{k}}\right \}\right ]
\end{eqnarray}
One may note that the constant $C_{20}$ which is proportional to the
mean pressure gradient remains arbitrary in the solution. In the
equations (1-2), if each term is averaged over an interval of time i.e
the period of oscillation, the solution of the present problem can be
derived from (8), (18), (20), (21), (30) and (33). The mean pressure
gradient may then be expressed as
\begin{eqnarray}
\overline{\frac{\partial p}{\partial x}}=\epsilon^2 \overline{\left(\frac{\partial p}{\partial x}\right)}_2 = \frac{\epsilon^2}{2e R}\phi_{20}^{\prime\prime\prime}-\frac{\epsilon^2}{kR}\phi_{20}^\prime+\frac{\epsilon^2}{4}i\alpha\left(\phi_1\phi_1^{\ast \prime\prime}- \phi_1^\ast \phi_1^{ \prime\prime}\right)+O(\epsilon^3)= \frac{\epsilon^2}{e R}C_{20}+O(\epsilon^3)
 \end{eqnarray}
As a natural consequence of (36), we can write
\begin{eqnarray}
  \overline{\left(\frac{\partial p}{\partial x}\right)}_2=  \frac{C_{20}}{e R}~~~~~~~~~~~~~~~~~~~~~~~~~~~~~~~~~~~~~~~~~~~~~~~~~~~~~~~~~~~~~~~~~~~~~~~~~~~~~~ 
 \end{eqnarray}
~~~~~It is now clear that $C_{20}$ is proportional to the time averaged pressure gradient arising out of the peristaltic motion of the fluid. Its value can be determined by employing the conditions for a given physiological problem. Noting that $\overline{\left(\frac{\partial p}{\partial x}\right)}_2$ is independent of the y-coordinate, the expression for the mean velocity in the axial direction can be put in the form
\begin{eqnarray}
\bar{u} = \frac{\epsilon^2}{2}\phi_{20}^{\prime} =\frac{\epsilon^2}{2}\left[f(y)+\frac{\left ( D-f(1)-sf^{\prime} (1) \right ) \cosh \frac{\sqrt{e}y}{\sqrt{k}}}{\cosh \frac{\sqrt{e}}{\sqrt{k}}+\frac{s\sqrt{e}}{\sqrt{k}} \sinh \frac{\sqrt{e}}{\sqrt{k}}}-\frac{2C_{20}k}{e} \left ( 1-\frac{\cosh \frac{\sqrt{e}y}{\sqrt{k}}}{ \cosh \frac{\sqrt{e}}{\sqrt{k}}+\frac{s\sqrt{e}}{\sqrt{k}}\sinh \frac{\sqrt{e}}{\sqrt{k}}} \right )\right]~~ 
\end{eqnarray}
Using (36), we can alternatively write  
\begin{eqnarray}
\bar{u} =\frac{\epsilon^2}{2}\left[f(y)+\frac{\left ( D-f(1)-sf^{\prime} (1) \right ) \cosh \frac{\sqrt{e}y}{\sqrt{k}}}{\cosh \frac{\sqrt{e}}{\sqrt{k}}+\frac{s\sqrt{e}}{\sqrt{k}}\sinh \frac{\sqrt{e}}{\sqrt{k}}}-2kR\overline{\left(\frac{\partial p}{\partial x}\right)}_2 \left ( 1-\frac{\cosh \frac{\sqrt{e}y}{\sqrt{k}}}{\cosh \frac{\sqrt{e}}{\sqrt{k}}+\frac{s\sqrt{e}}{\sqrt{k}}\sinh \frac{\sqrt{e}}{\sqrt{k}}} \right )\right]~~~~~~ 
\end{eqnarray}

\begin{figure}
 \includegraphics[width=3.3in,height=2.4in]{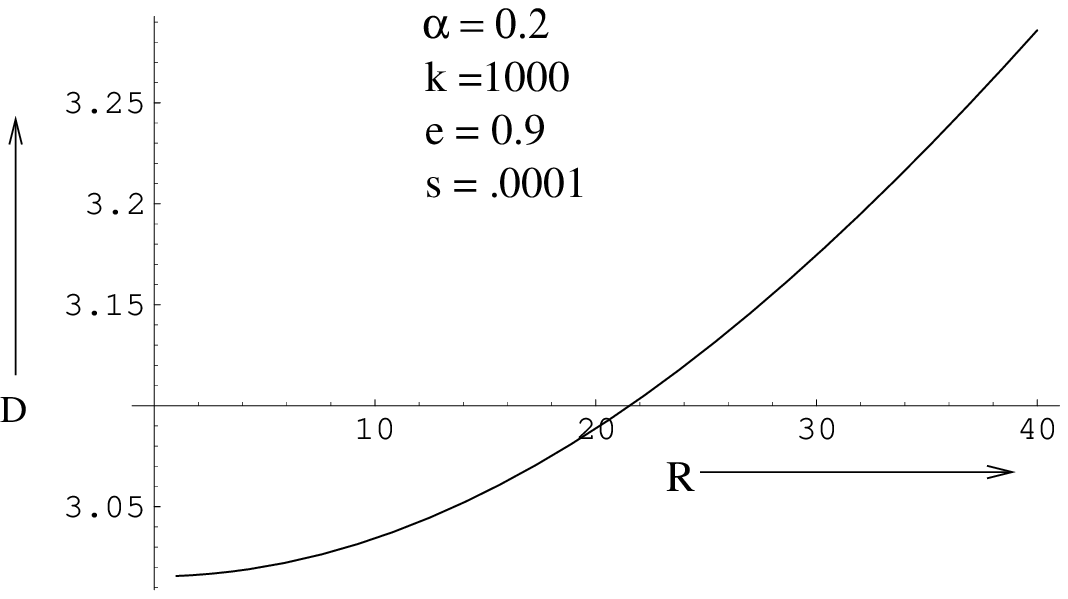}\hspace{.2cm}\includegraphics[width=3.3in,height=2.4in]{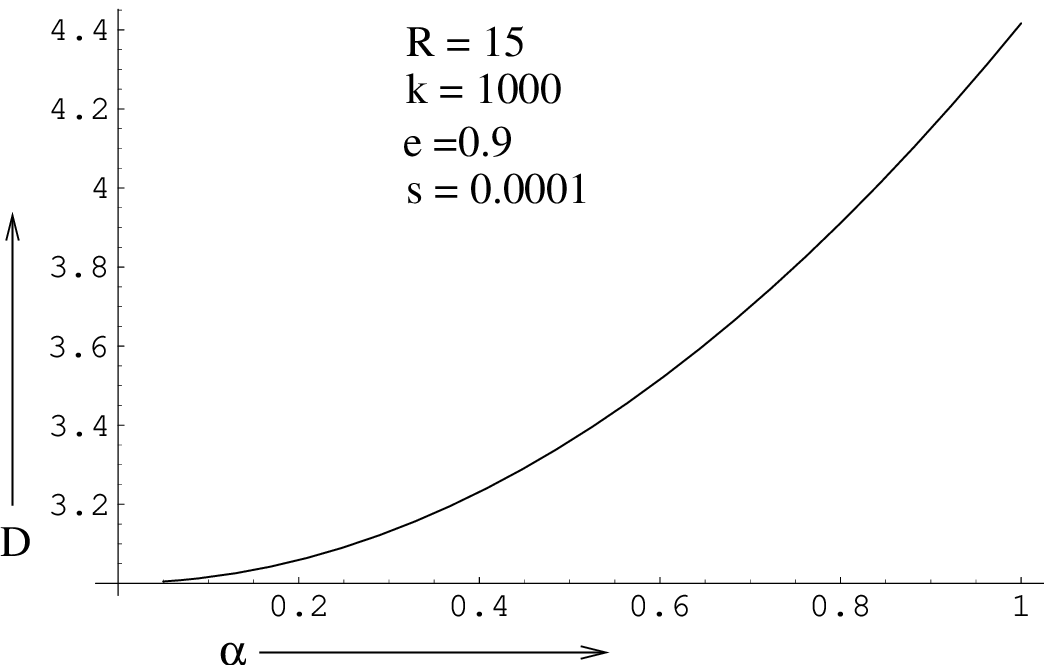} \\$~~~~~~~~~~~~~~~~~~~~~~~~~~~~~~~~~~(a)~~~~~~~~~~~~~~~~~~~~~~~~~~~~~~~~~~~~~~~~~~~~~~~~~~~~~~~~~~~~~~~~~~~~~~~~~~~~~~~~~~~~~~(b)~~~~~~~~~$\\  
 \includegraphics[width=3.3in,height=2.4in]{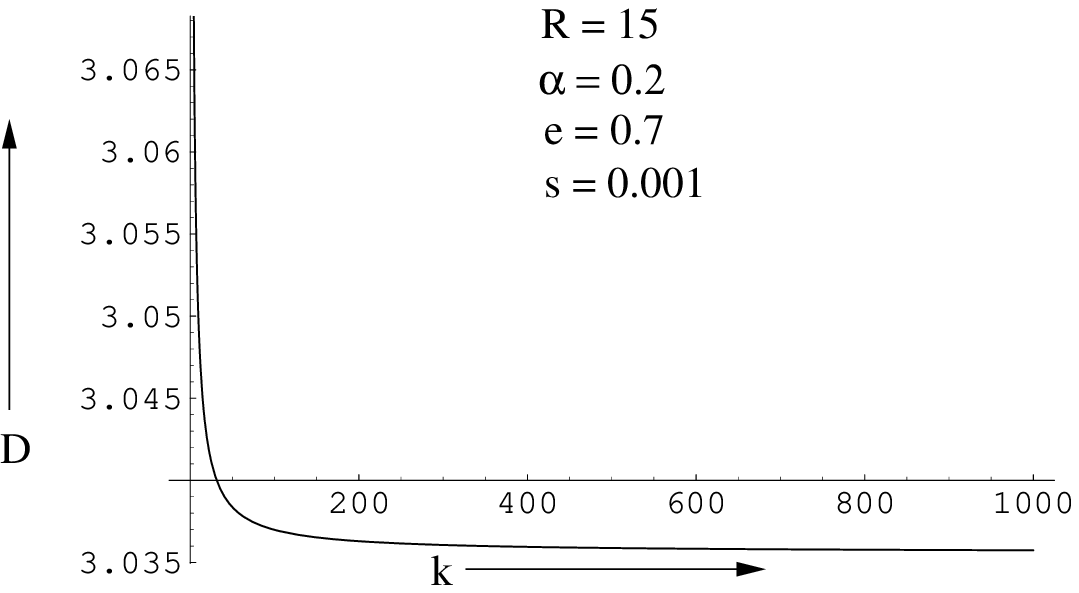}\hspace{.2cm}\includegraphics[width=3.3in,height=2.4in]{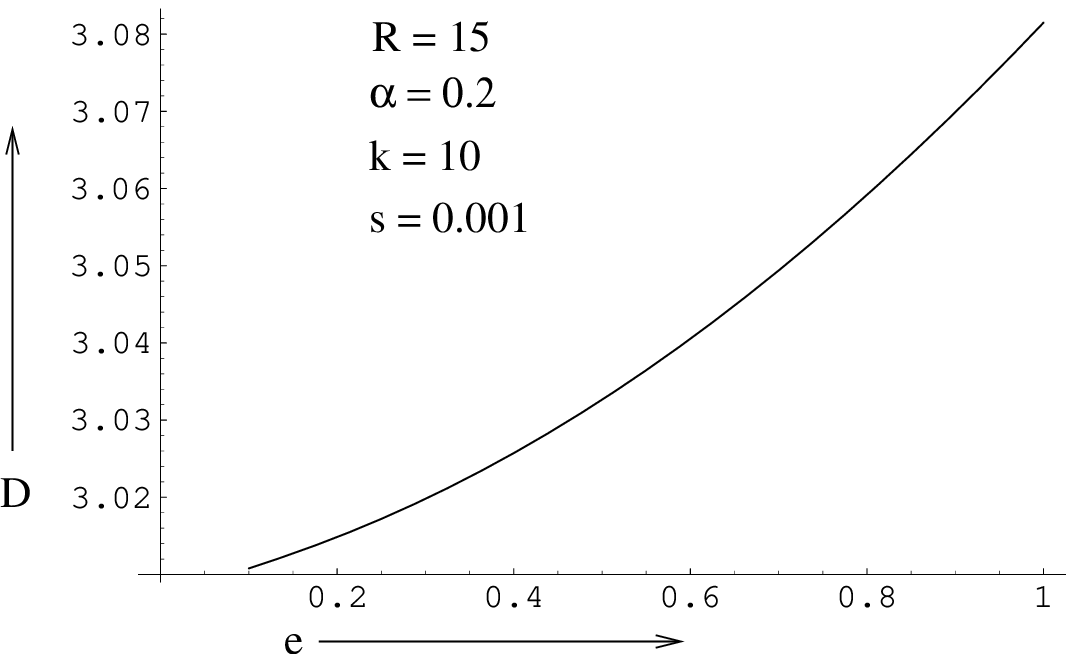} \\$~~~~~~~~~~~~~~~~~~~~~~~~~~~~~~~~~~(c)~~~~~~~~~~~~~~~~~~~~~~~~~~~~~~~~~~~~~~~~~~~~~~~~~~~~~~~~~~~~~~~~~~~~~~~~~~~~~~~~~~~~~~(d)~~~~~~~~~$\\ 
      \begin{center}
\includegraphics[width=3.5in,height=2.4in]{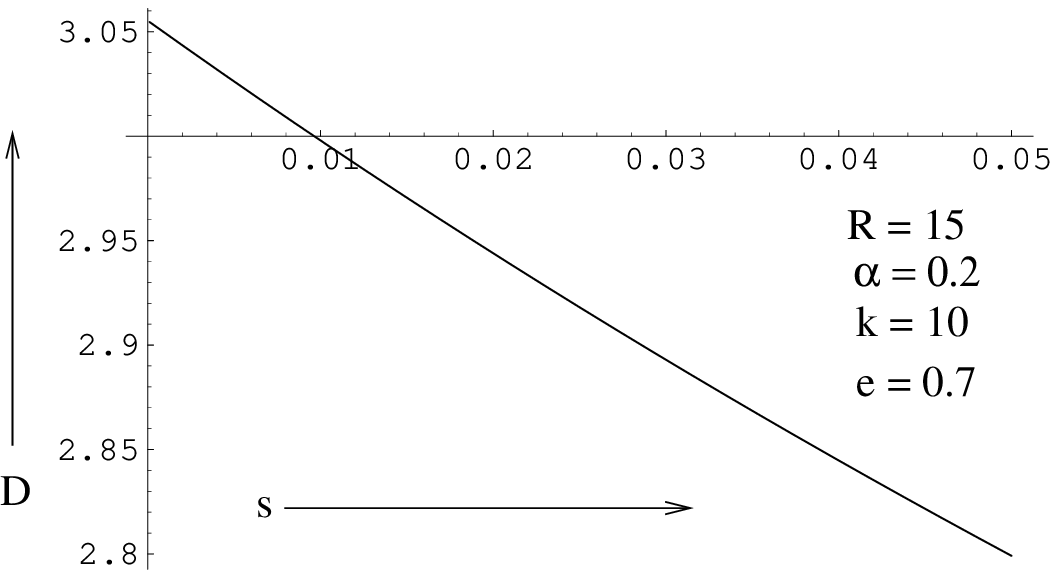}\\(e)
\end{center}
Fig. 2: Variation of D with R, $\alpha$, Darcy number, porosity and slip parameter\\ 
\end{figure}

For a problem such as the one under our present considerations, it is
worthwhile to determine the critical reflux condition. This is the
condition that determines the location of the point on the central
line (y=0) at which the mean velocity $\bar{u}(y)$ vanishes. Using
(39), the  critical reflux condition for the problem is derived in the form
\begin{eqnarray}
\overline{\left(\frac{\partial p}{\partial x}\right)}_{2~critical~reflux}=\frac{1}{2kR}\left[\frac{f(0)\left( \cosh \frac{\sqrt{e}}{\sqrt{k}}+\frac{\sqrt{e}}{\sqrt{k}} \sinh \frac{\sqrt{e}}{\sqrt{k}}\right)+ D-f(1)-sf^{\prime}(1)}{ \cosh \frac{\sqrt{e}}{\sqrt{k}}+\frac{s\sqrt{e}}{\sqrt{k}}\sinh \frac{\sqrt{e}}{\sqrt{k}}-1 }\right]~~~~~~~~~~~~~~~~~~~~~~ 
\end{eqnarray}
Use of this critical value enables us to study the mean velocity
distribution $\bar{u}(y)$ at the critical reflux condition. 
 
\begin{figure}
 \includegraphics[width=3.5in,height=2.6in]{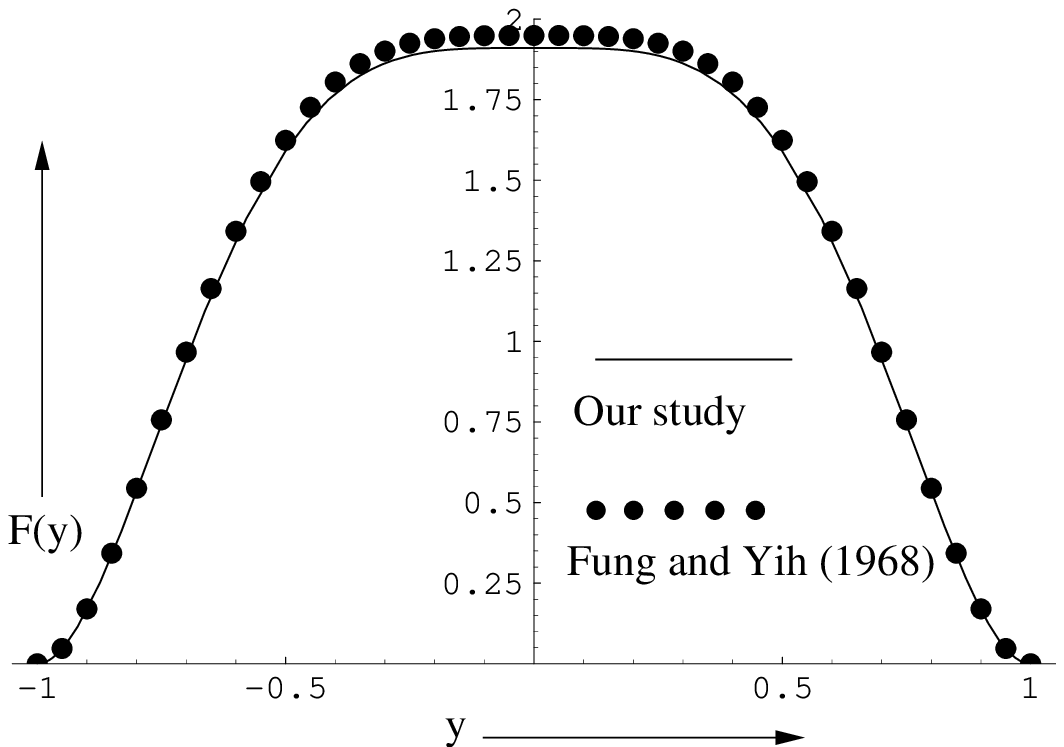}\includegraphics[width=3.5in,height=2.6in]{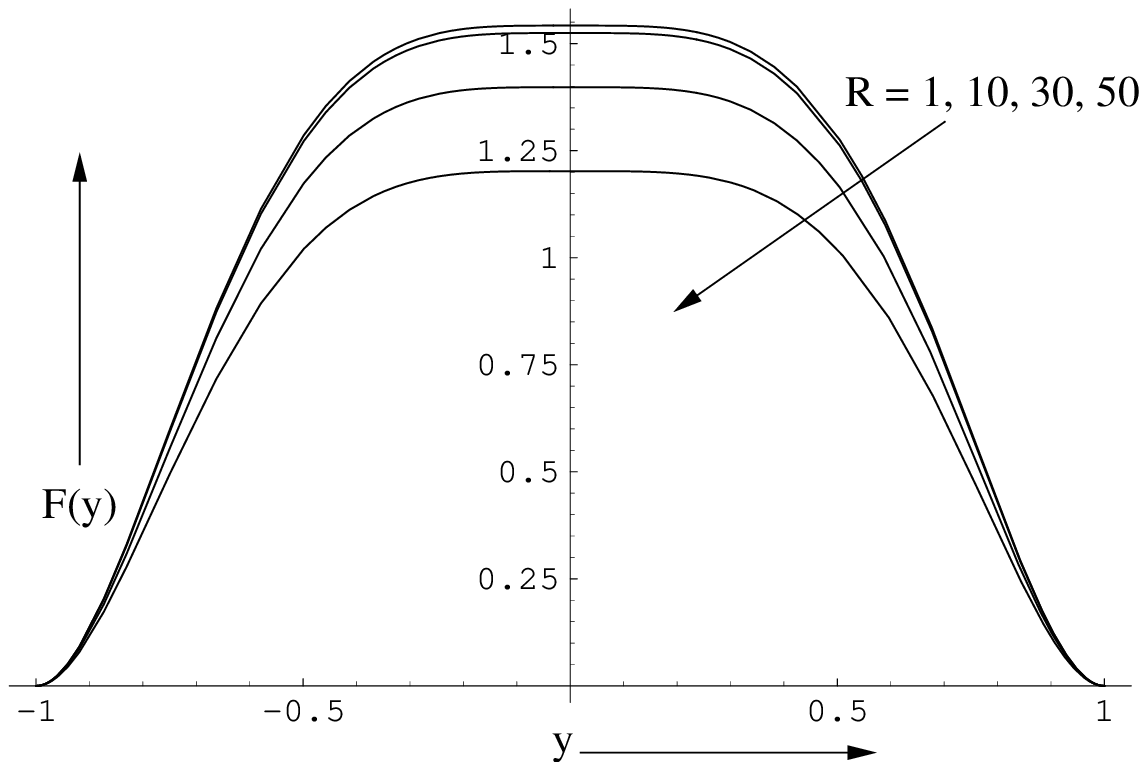}
 \\$~~~~~~~~~~~~~~~~~~~~~~~~~~~~~~~~~~Fig. 3~~~~~~~~~~~~~~~~~~~~~~~~~~~~~~~~~~~~~~~~~~~~~~~~~~~~~~~~~~~~~~~~~~~~~~~~~~~~~~~~~~~~~~Fig. 4~~~~~~~~~$\\ Fig. 3-4:
 Mean velocity perturbation function F(y), Fig3: Comparison of our
 result with Fung and Yih \cite{r2} for R=1, $\alpha=$0.4, k=10000,
 e=0.99, s=0.00001; Fig4: Distribution of F(y) for different values of
 R with $\alpha=$0.25, k=1000, e=0.9, s=0.00001 \\
\end{figure}    
\section{Numerical Results and Discussion}

In the last section, we have presented the analytical expressions of
the velocity, time-averaged velocity and critical pressure for reflux
for the problem under consideration. In this section, we intend to
investigate the problem and present the computational results for the
said quantities graphically. Computation has been performed by making
use of available experimental and theoretical data of bile. The
different parameters involved in our forgoing theoretical analysis are
Reynolds number R, pressure gradient, porosity parameter e, Darcy
number k, slip parameter, amplitude ratio and wave number. Computation
of the time-averaged velocity and critical pressure for reflux has
been carried out by extensive use of the software Mathematica.
\begin{figure}
 \includegraphics[width=3.5in,height=2.6in]{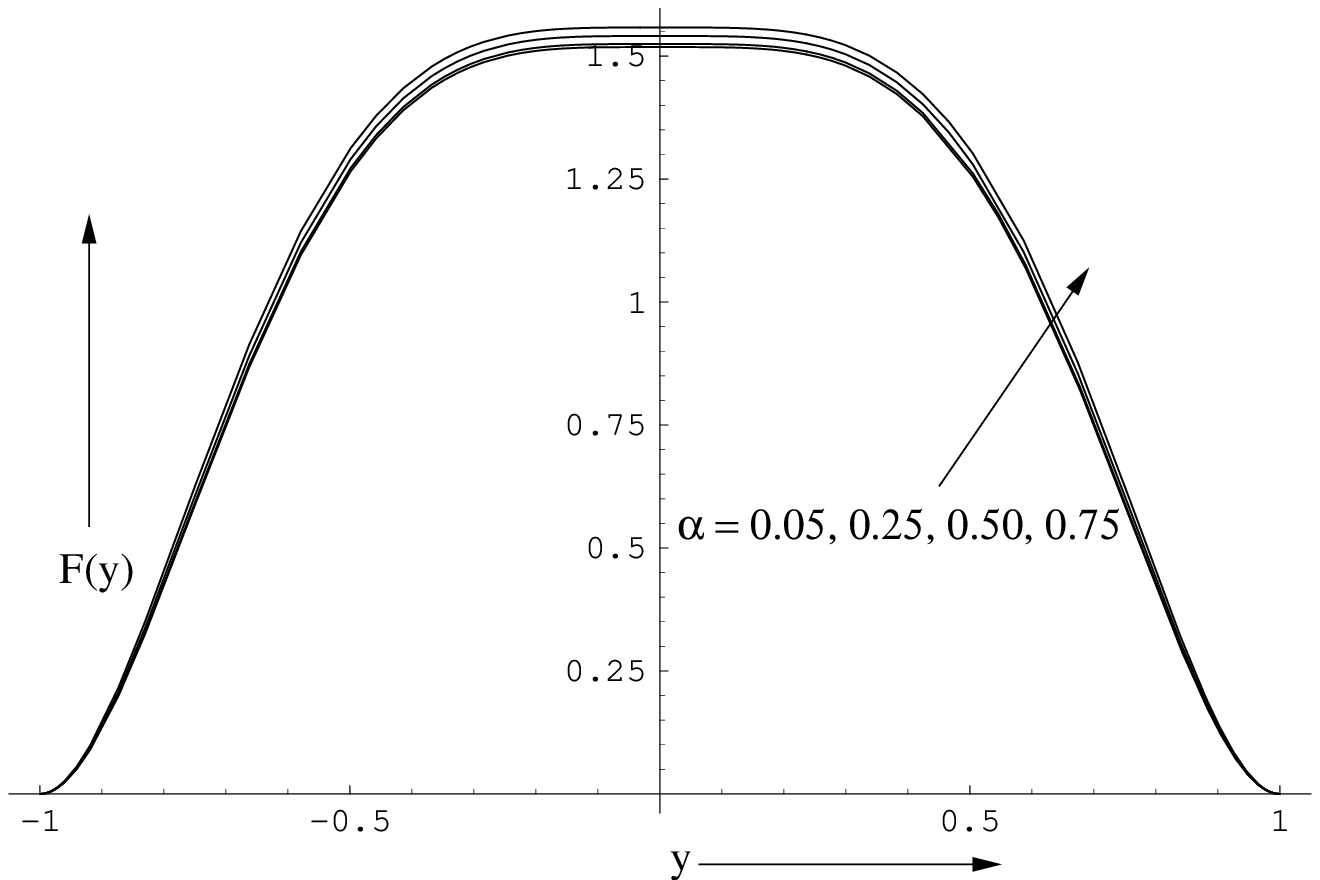}\includegraphics[width=3.5in,height=2.6in]{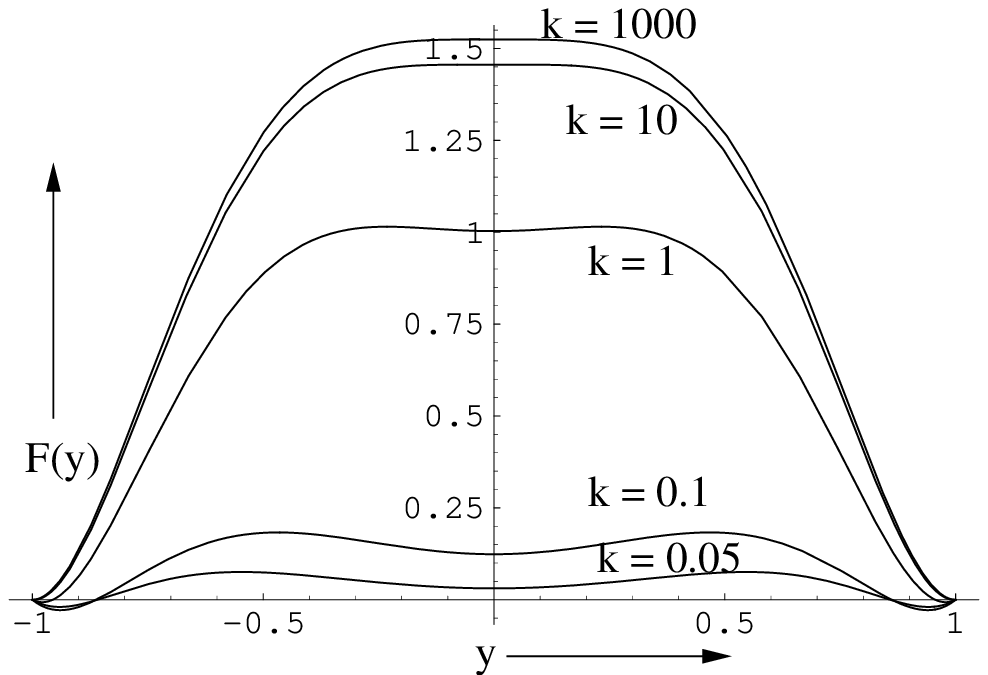} \\$~~~~~~~~~~~~~~~~~~~~~~~~~~~~~Fig. 5~~~~~~~~~~~~~~~~~~~~~~~~~~~~~~~~~~~~~~~~~~~~~~~~~~~~~~~~~~~~~~~~~~~~~~~~~~~~~~~~~~~~~~~Fig. 6 ~~~~~~~~~~~~~~~~~~$\\ 
\\ 
 \includegraphics[width=3.5in,height=2.6in]{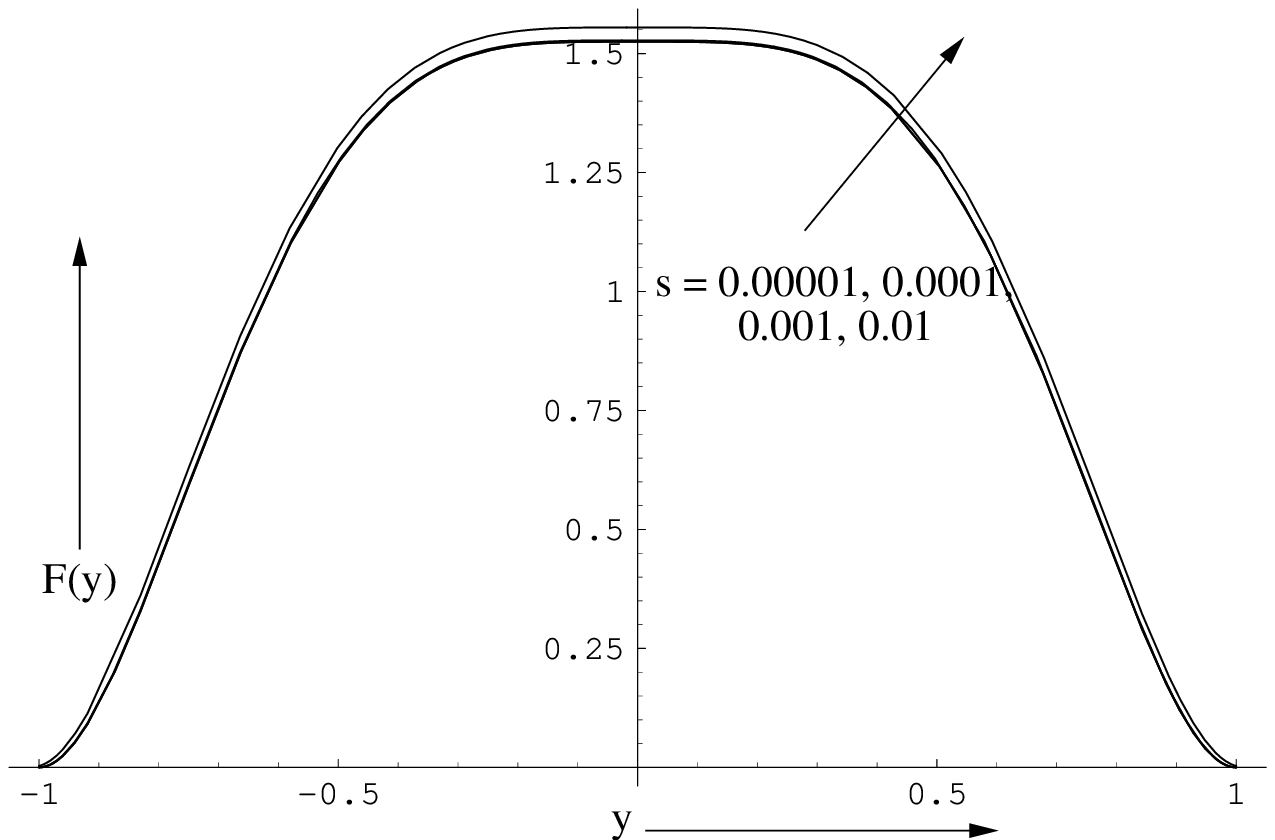}\includegraphics[width=3.5in,height=2.6in]{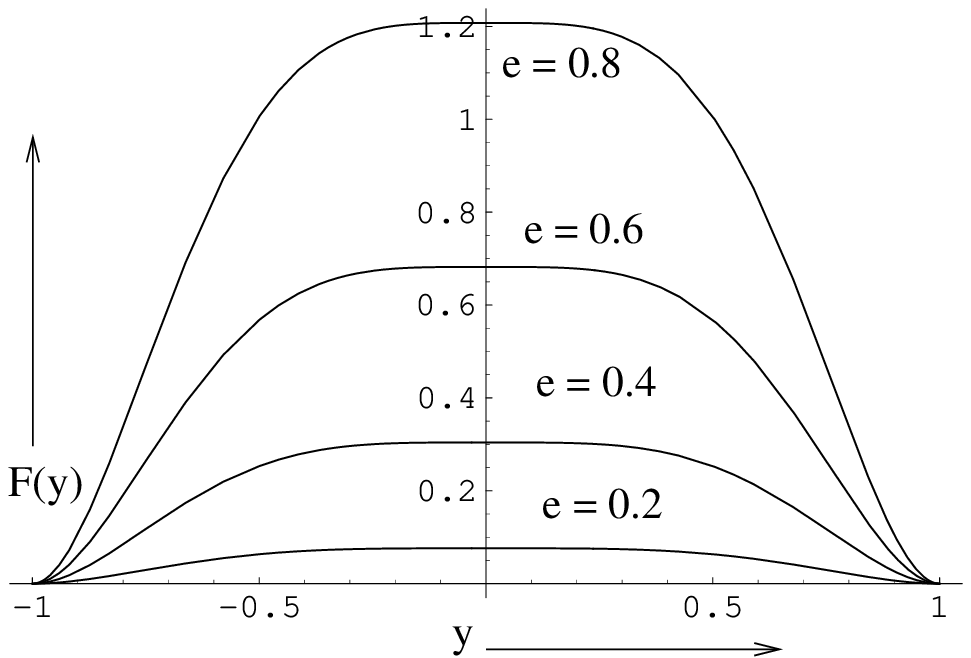} \\$~~~~~~~~~~~~~~~~~~~~~~~~~~~~~Fig. 7~~~~~~~~~~~~~~~~~~~~~~~~~~~~~~~~~~~~~~~~~~~~~~~~~~~~~~~~~~~~~~~~~~~~~~~~~~~~~~~~~~~~~~~Fig. 8 ~~~~~~~~~~~~~~~~~~$\\ 
\\      
Figs. 5-8: Mean velocity perturbation function F(y) (Fig5.) for different values of $\alpha$ with R=10, k=1000, e=0.9, s=0.00001; (Fig6.) for different values of k with R=10, $\alpha$=0.25, e=0.9, s=0.00001; (Fig7.) for different values of e with R=10, $\alpha$=0.25, k=1000, s=0.00001; (Fig8.) for different values of s with R=10, $\alpha$=0.25, e=0.9, k=1000\\ 
\end{figure}
\begin{figure}
 \includegraphics[width=3.5in,height=2.4in]{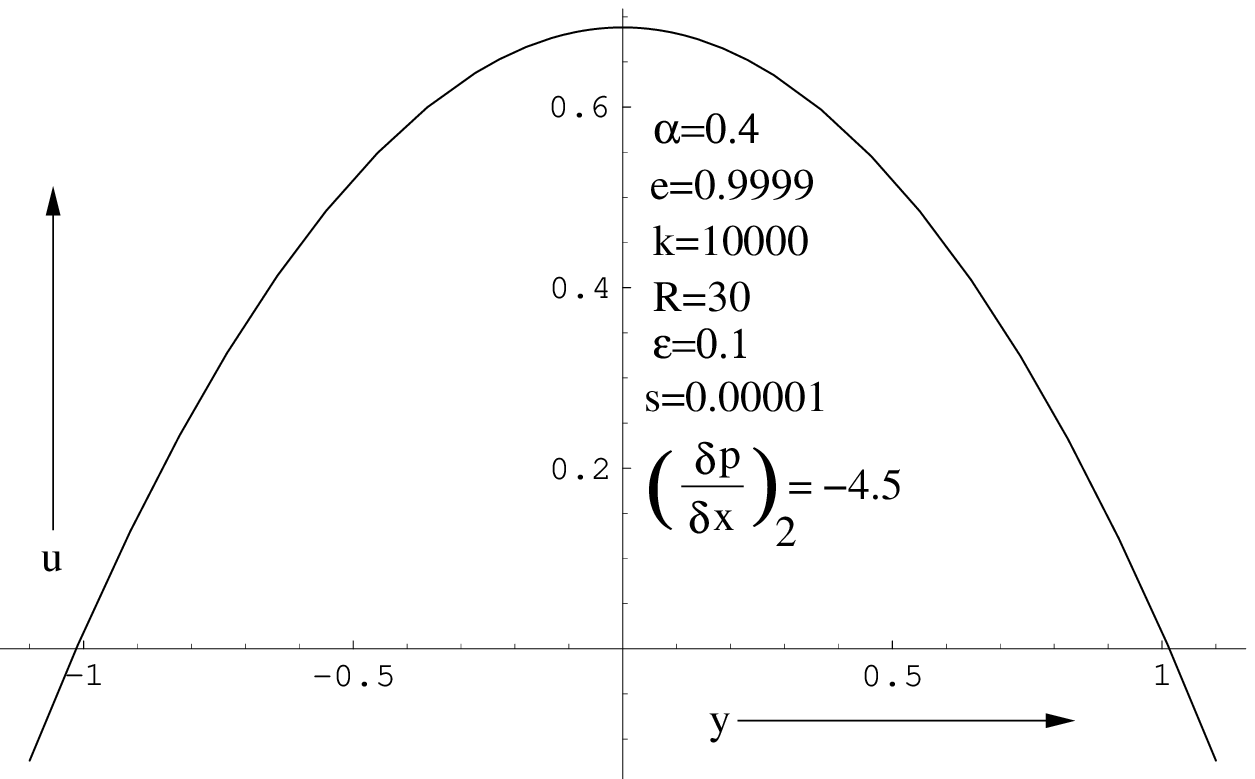}\includegraphics[width=3.5in,height=2.4in]{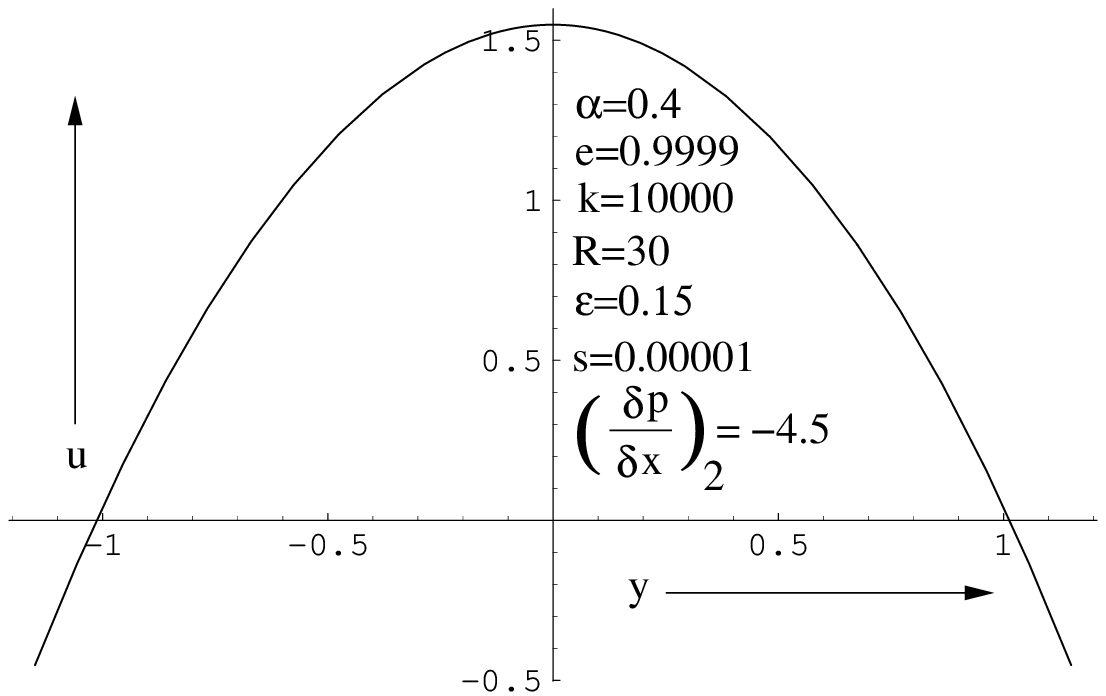}\\$~~~~~~~~~~~~~~~~~~~~~~~~~~~~~~~~~Fig. 9.1~~~~~~~~~~~~~~~~~~~~~~~~~~~~~~~~~~~~~~~~~~~~~~~~~~~~~~~~~~~~~~~~~~~~~~~~~~~~~~~~~~~~~~Fig 9.2~~~~~~~~~$\\ 
 \includegraphics[width=3.5in,height=2.4in]{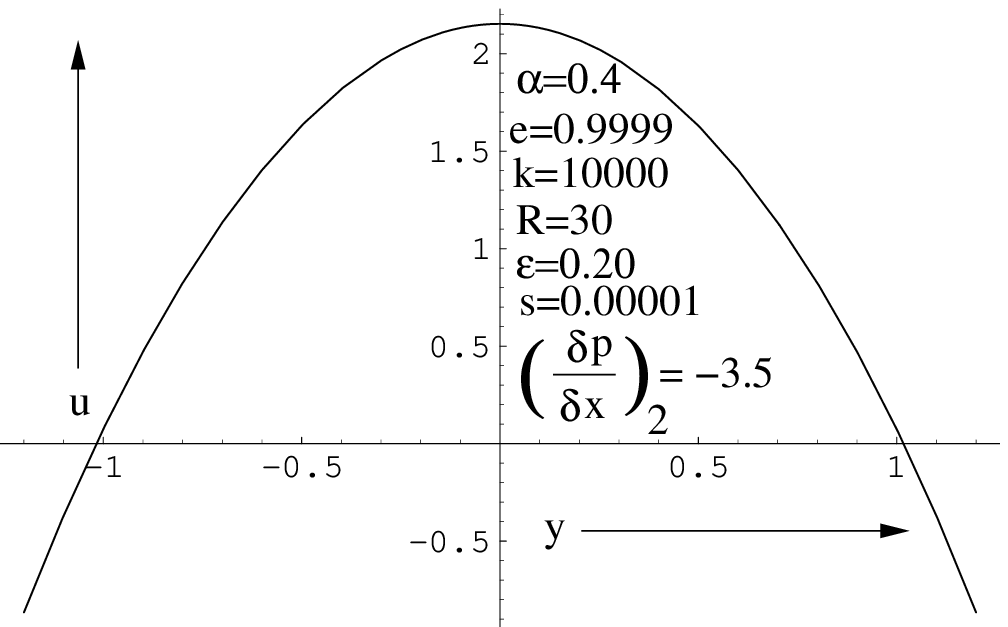}\includegraphics[width=3.5in,height=2.4in]{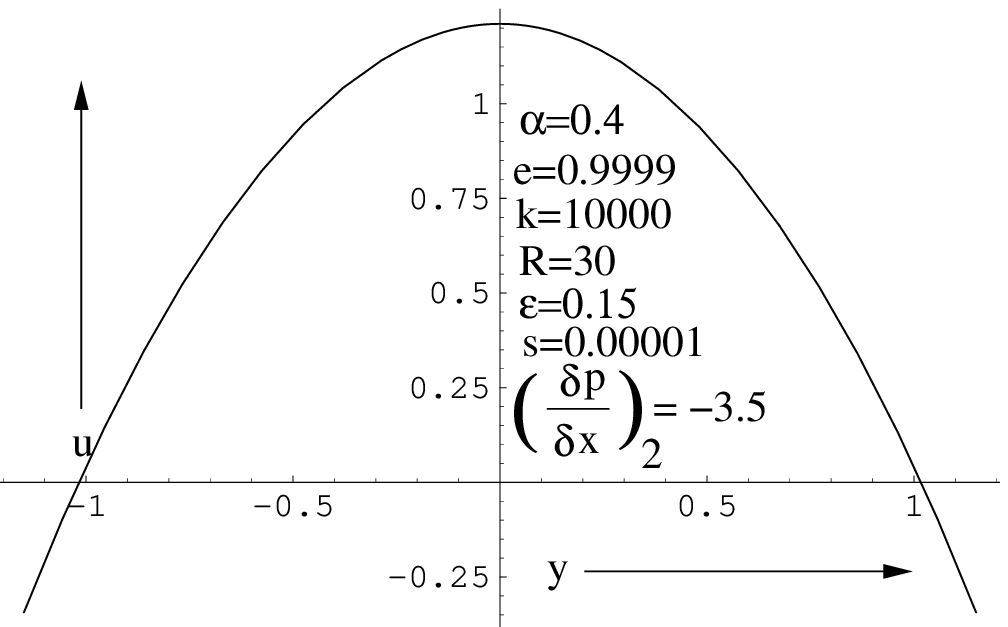}\\$~~~~~~~~~~~~~~~~~~~~~~~~~~~~~~Fig. 9.3~~~~~~~~~~~~~~~~~~~~~~~~~~~~~~~~~~~~~~~~~~~~~~~~~~~~~~~~~~~~~~~~~~~~~~~~~~~~~~~~~~~~~~~~~~~Fig. 9.4 ~~~~~~~~~~~~~~$\\ 
\includegraphics[width=3.5in,height=2.4in]{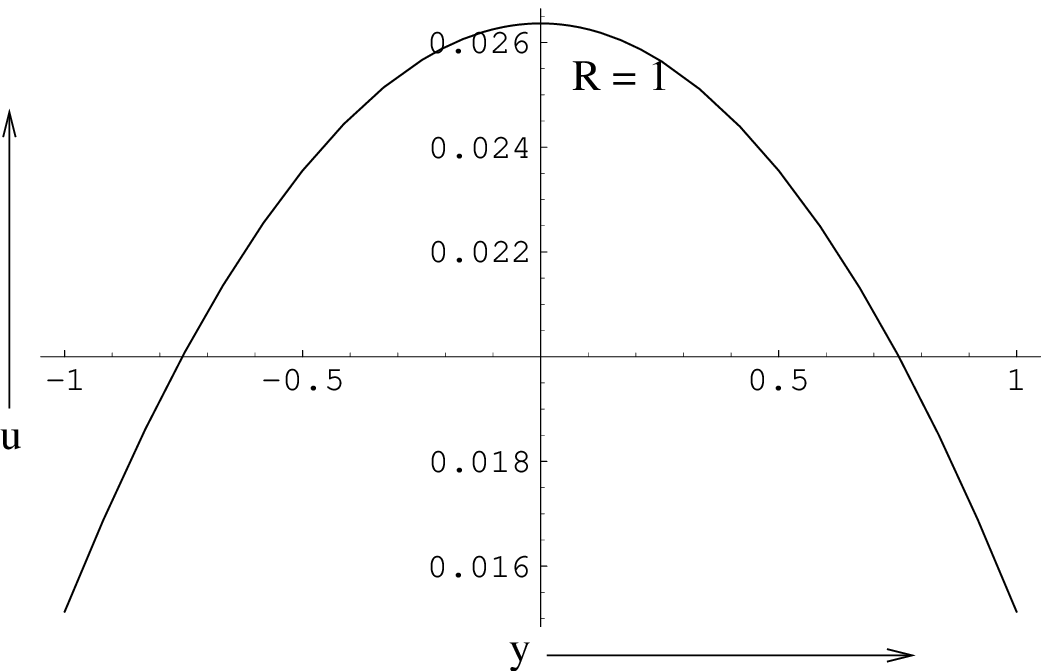}\includegraphics[width=3.5in,height=2.4in]{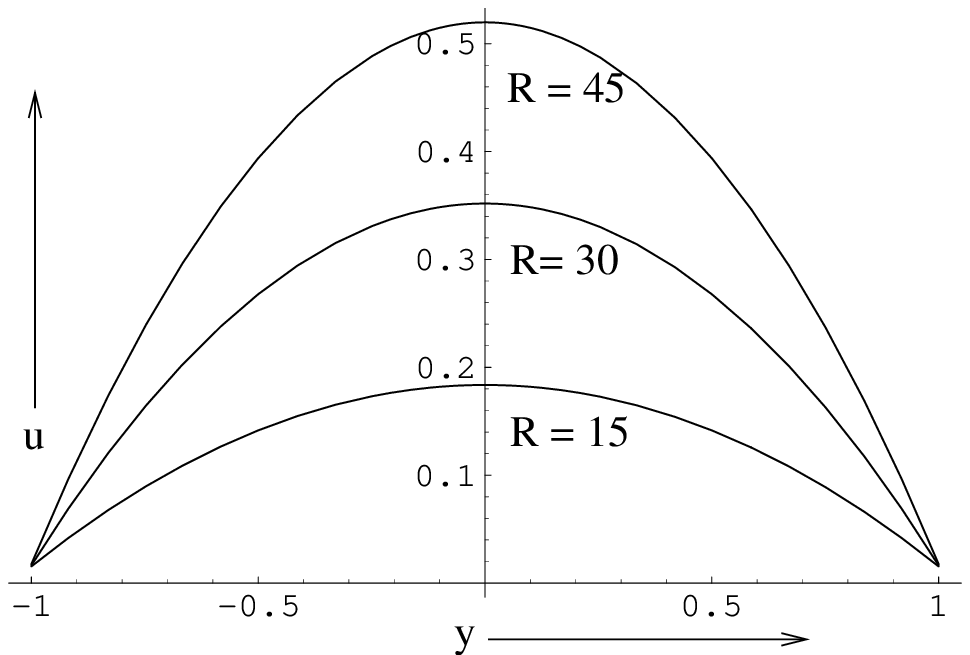}\\$~~~~~~~~~~~~~~~~~~~~~~~~~~~~~~~~~Fig. 9.5~~~~~~~~~~~~~~~~~~~~~~~~~~~~~~~~~~~~~~~~~~~~~~~~~~~~~~~~~~~~~~~~~~~~~~~~~~~~~~~~~~~~~~Fig. 9.6~~~~~~~~~$\\ 
Figs. 9:  Time-averaged mean axial velocity profiles for different values of  R when $\alpha$=0.25, e=0.9, $\epsilon$=0.1, $\overline{\left(\frac{\partial p}{\partial x}\right)}_2=-2.5$, k=1000, s=0.0001
\\
\end{figure}
\begin{figure}
 \includegraphics[width=3.7in,height=2.8in]{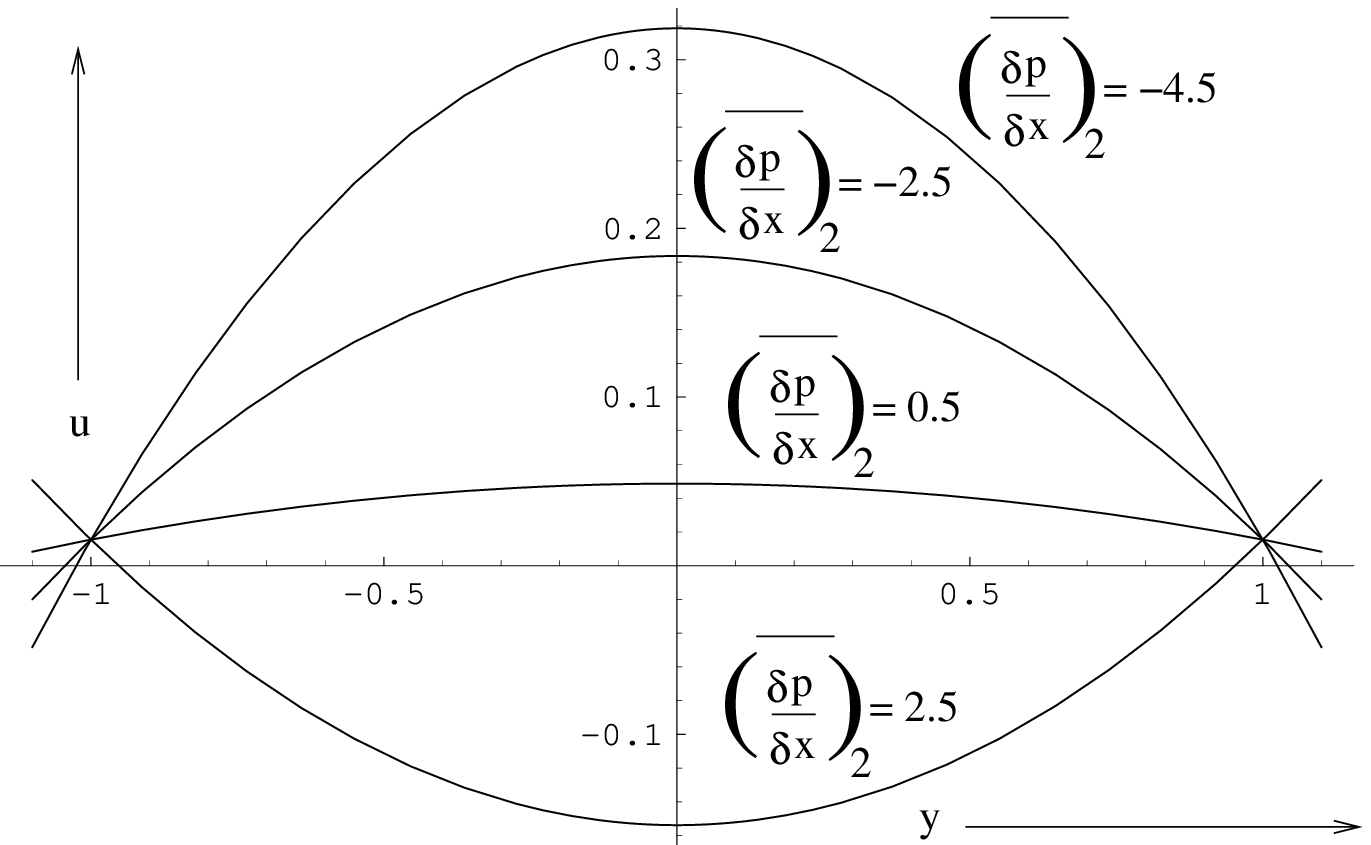}\includegraphics[width=3.7in,height=2.8in]{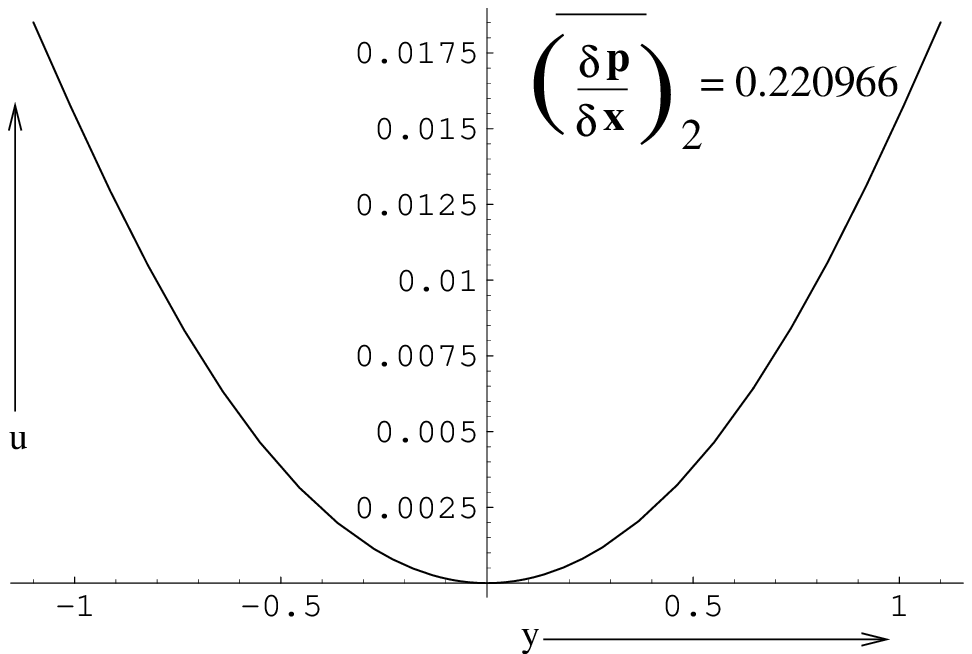}                    
\\$~~~~~~~~~~~~~~~~~~~~~~~~~~~~~Fig. 10.1~~~~~~~~~~~~~~~~~~~~~~~~~~~~~~~~~~~~~~~~~~~~~~~~~~~~~~~~~~~~~~~~~~~~~~~~~~~~~~~~~~~~~~~~~Fig. 10.2 ~~~~~~~~~~~~~~$\\ 
Figs. 10: Time-averaged mean axial velocity profiles for different values of $\overline{\left(\frac{\partial p}{\partial x}\right)}_2$ when R=15, $\alpha$=0.25, e=0.9, $\epsilon$=0.1, k=1000, s=0.0001
\end{figure}\\

\subsection{ Average Velocity in Bile Transport}

Numerical computation based on equation (39) reveals that the averaged axial velocity of bile (if treated as a Newtonian fluid) flowing in a porous medium is dominated by the constant \begin{eqnarray*}  \frac{D}{ \cosh \frac{\sqrt{e}}{\sqrt{k}}+\frac{s\sqrt{e}}{\sqrt{k}} \sinh \frac{\sqrt{e}}{\sqrt{k}}}~and~ the~ parabolic~ distribution~ term~2kR\overline{\left(\frac{\partial p}{\partial x}\right)}_2 \left ( 1-\frac{\cosh \frac{\sqrt{e}y}{\sqrt{k}}}{ \cosh \frac{\sqrt{e}}{\sqrt{k}}+\frac{s\sqrt{e}}{\sqrt{k}}\sinh \frac{\sqrt{e}}{\sqrt{k}}} \right )\end{eqnarray*}
 In addition to these two terms, there is an expression 
\begin{eqnarray}
G(y)=f(y)-\frac{\left ( f(1)+sf^{\prime} (1) \right ) \cosh \frac{\sqrt{e}y}{\sqrt{k}}}{ \cosh \frac{\sqrt{e}}{\sqrt{k}}+\frac{s\sqrt{e}}{\sqrt{k}} \sinh \frac{\sqrt{e}}{\sqrt{k}}}
\end{eqnarray}
that represents the perturbation of the velocity across the
channel. Its distribution controls the direction of peristaltic mean
flow across the cross-section of the bile duct. It is evident from
(33) and (31) that the constant D that gives the value of
$\phi_{20}^{\prime}$ at the boundary depends on Reynolds number,
porosity, Darcy number,slip parameter and wave number. It arises from
the expression of the radial component of the first-order axial
velocity gradient and is connected with the offset boundary condition
(27). It may be mentioned here that the no-slip/velocity-slip applies
to the wavy wall and not to the mean position of the
wall. Figs. 2(a-e) give the variation of D with the parameters
Reynolds number, porosity, Darcy number, wave number and
velocity-slip.
\begin{eqnarray*}Furthermore,~the~ parabolic~ mean~velocity~ distribution~term~ -2kC_{20}\left ( 1-\frac{\cosh \frac{\sqrt{e}y}{\sqrt{k}}}{ \cosh \frac{\sqrt{e}}{\sqrt{k}}+\frac{s\sqrt{e}}{\sqrt{k}}\sinh \frac{\sqrt{e}}{\sqrt{k}}} \right )\\ arises~ out~ of~the~time- averaged~ second~order~ pressure~ gradient ~\overline{\left(\frac{\partial p}{\partial x}\right)}_2=\frac{C_{20}}{eR}.~~~~~~~~~
\end{eqnarray*}
We define F(y)= $-\frac{200}{\alpha^2 R^2}G(y)$, G(y) being given by
(41). Fig.3 depicts the nature of the function F(y). With an aim to
validate our results, we have presented in the same figure the results
reported by Fung and Yih \cite{r2} who did not account for the Darcy
number, porosity and velocity-slip. It is needless to mention that the
results presented in this figure on the basis of our study correspond
to the case when Darcy number, porosity and velocity-slip are
neglected (for the sake of meaningful comparison/validation). It may
be noted that the results of both the studies have a very close
resemblance. In Fig. 4 we have presented the nature of variation of
F(y) for different values of the Reynolds number for a particular set
of values of $\alpha$, k, e and s. This figure shows that F(y)
decreases as the Reynolds number R increases. This observation is in
conformity to that reported in \cite{r2}. It may be noted from Figs. 5
and 7 that in the ranges of values for $\alpha$ (wave number) and s
(slip parameter) considered here, the change in the function F(y) is
not very much significant.
\begin{figure}
 \includegraphics[width=3.7in,height=2.8in]{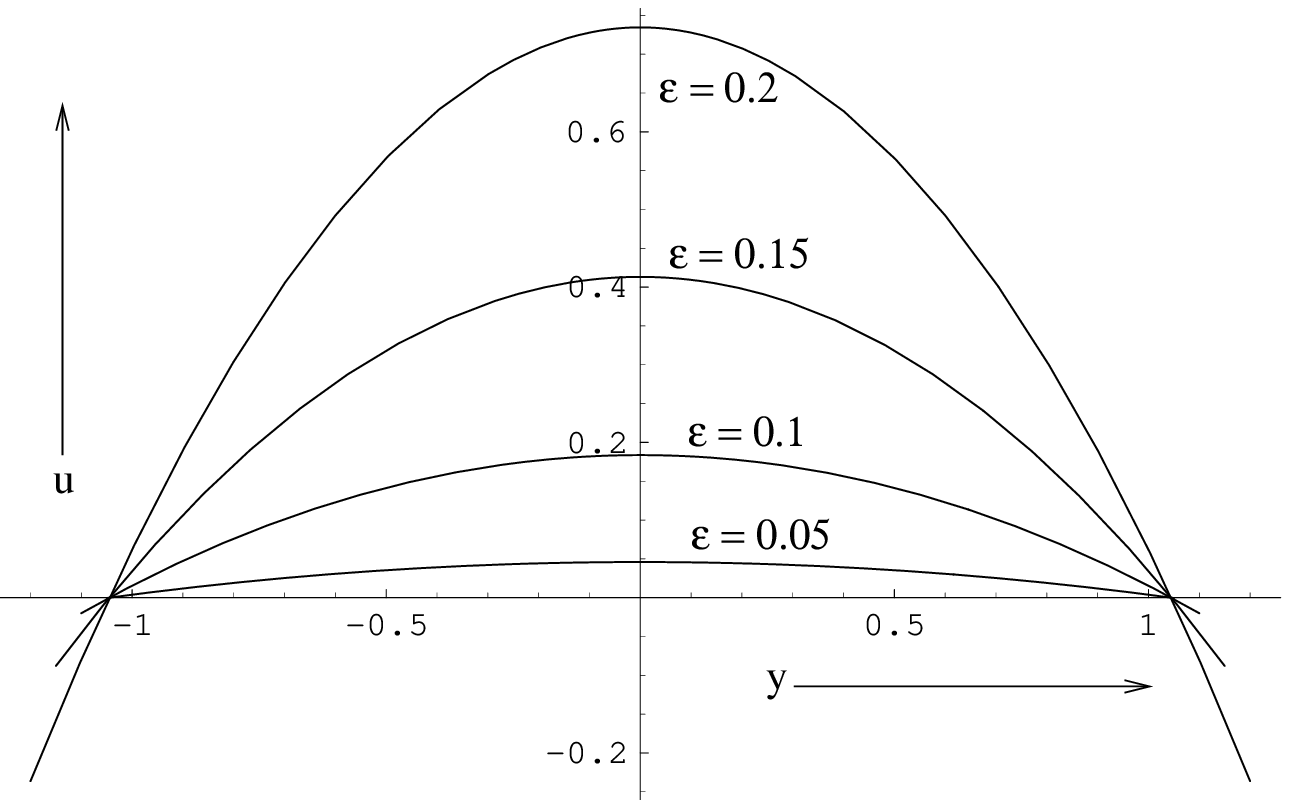}\includegraphics[width=3.7in,height=2.8in]{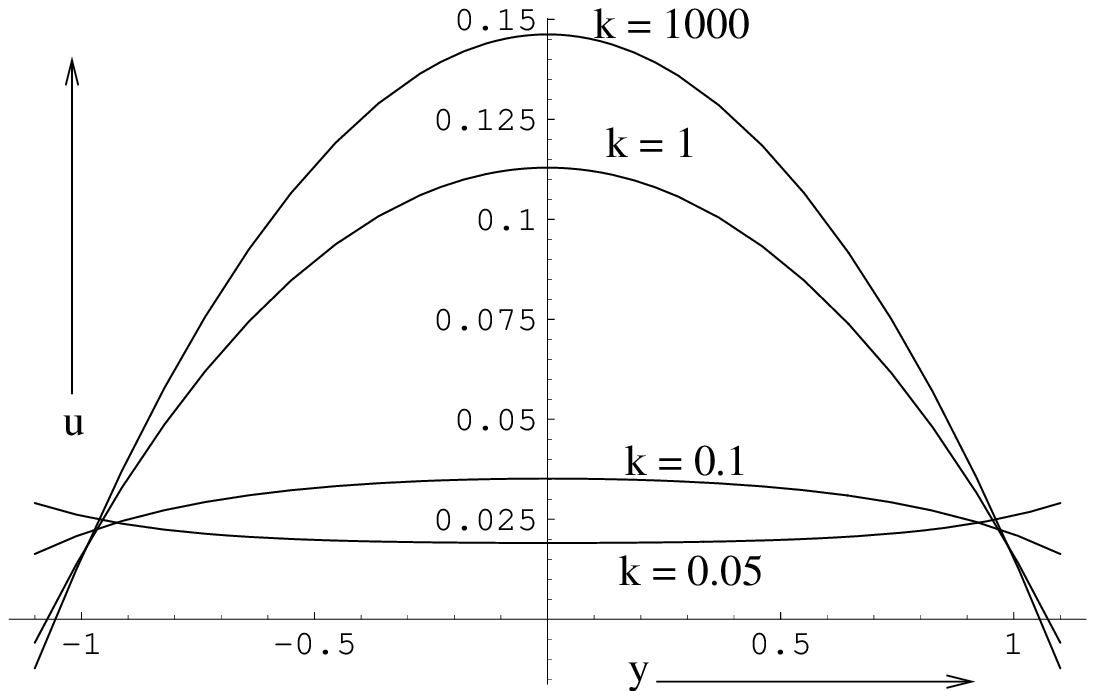}\\$~~~~~~~~~~~~~~~~~~~~~~~~~~~~~~~~~~~~~Fig. 11~~~~~~~~~~~~~~~~~~~~~~~~~~~~~~~~~~~~~~~~~~~~~~~~~~~~~~~~~~~~~~~~~~~~~~~~~~~~~~~~~~~~~~~~~~Fig. 12~~~~~~~~~$\\ \\ Figs. 11-12:
 Time averaged mean axial velocity profiles, Fig. 11: for different
 values of $\epsilon$ when R=15, $\alpha$=0.25, e=0.9,
 $\overline{\left(\frac{\partial p}{\partial x}\right)}_2=-2.5$,
 k=1000, s=0.0001; Fig. 12: for different values of k when R=15,
 $\overline{\left(\frac{\partial p}{\partial x}\right)}_2$=-2.5,
 $\alpha$=0.25, e=0.9, $\epsilon$=0.1, s=0.0001
\end{figure}\\
Figs. 6 and 8 reveal that with the increase in the Darcy number k and
the porosity parameter e, the perturbation term F(y) increases. The
results correspond to the case of bile containing stones. Figs. 9-15
illustrate the individual contribution of Reynolds number, pressure
gradient, amplitude ratio, Darcy number, the porosity parameter, slip
parameter and wave number on the time-averaged mean axial velocity
profile. It is found that the mean axial velocity increases with an
increase in Reynolds number (See Figs. 9). It is important to note
(cf. Figs. 10) that reflux occurs in the central region when pressure
gradient attains a certain critical value 0.220966. Fig. 11
illustrates that the bile flow is strongly dependent on the amplitude
ratio of the wave. It reveals that bile velocity is enhanced with the
increase in amplitude ratio and that flow reversal takes place near
the boundary for higher values of the amplitude ratio.  These results
are in conformity (cf. Figs. 9.1-4) to those of an experimental study
conducted by Sugita et al. \cite{r35}. They investigated pseudolesion
of the bile duct caused by flow effect and observed that at
1.0mm/sec. intervals flow speed of bile ranges between 1.0 and 20
mm/sec. approximately. It may be noted from Fig. 12 that although in
the central region of the channel, the velocity reduces with the
decrease in the Darcy number, the velocity profiles are quite
different from the familiar Poiseuille profile. This shows that the
bile velocity decreases as the number of stones increases.

\begin{figure}
 \includegraphics[width=3.6in,height=2.8in]{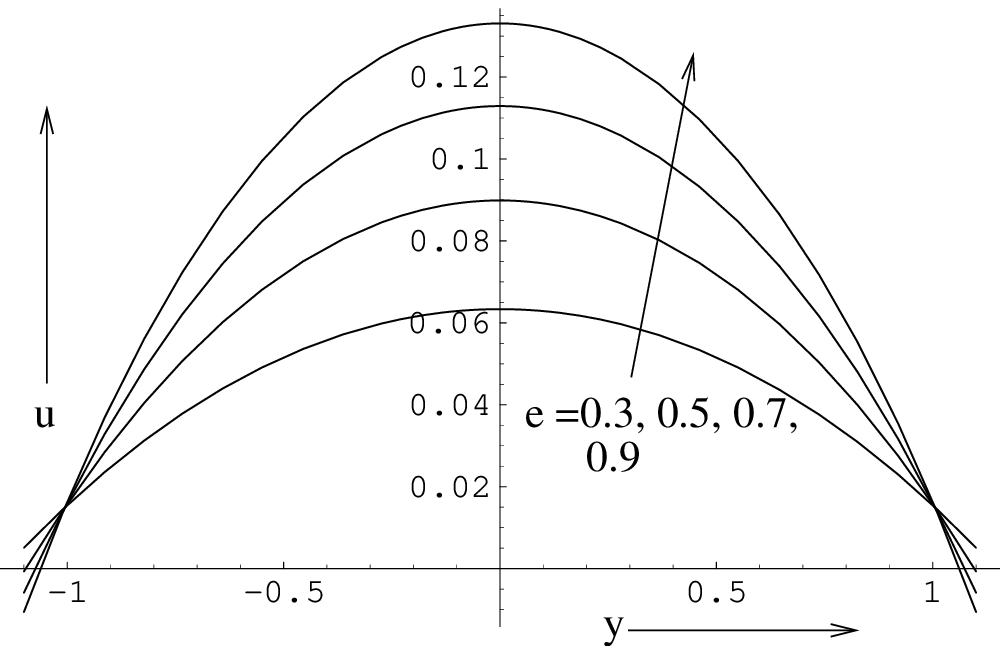}\includegraphics[width=3.6in,height=2.8in]{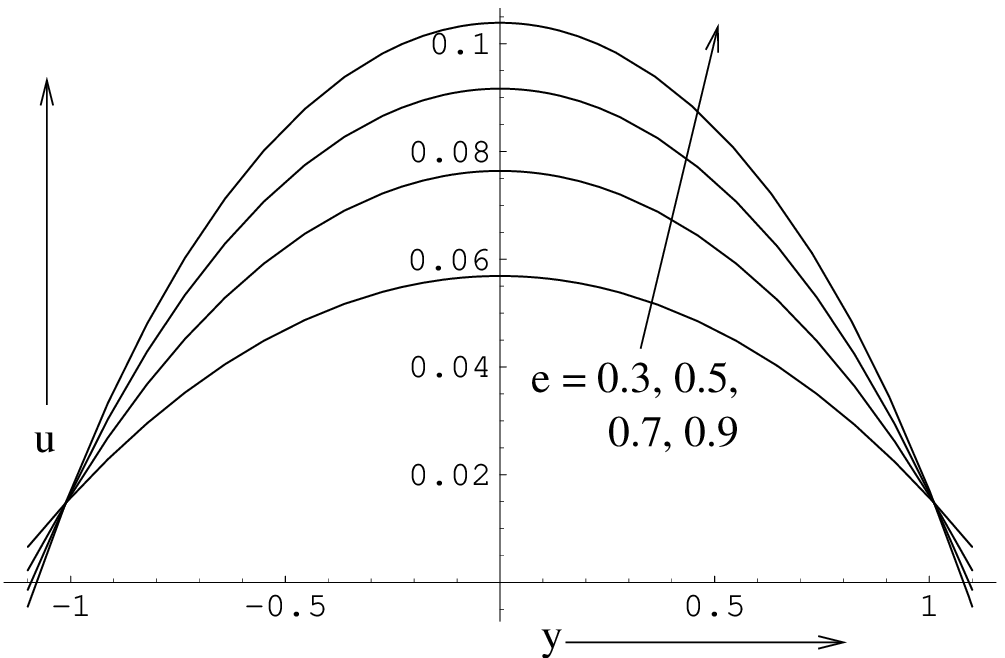}\\$~~~~~~~~~~~~~~~~~~~~~~~~~~~~~~~~~~~~~(a)~~~~~~~~~~~~~~~~~~~~~~~~~~~~~~~~~~~~~~~~~~~~~~~~~~~~~~~~~~~~~~~~~~~~~~~~~~~~~~~~~~~~~~~~~~(b)~~~~~~~~~$\\ 
\\
 \includegraphics[width=3.7in,height=2.8in]{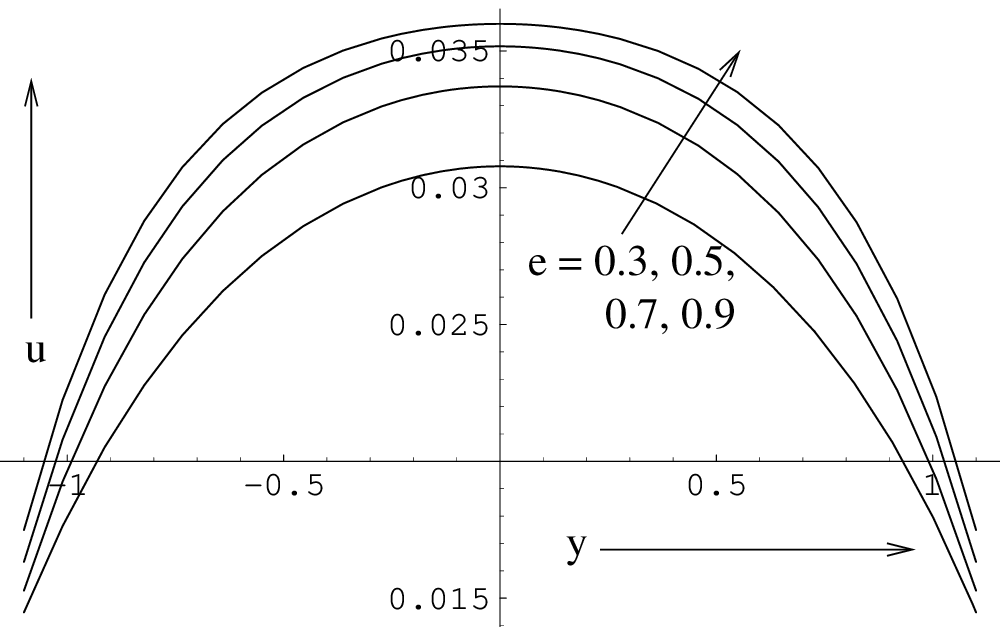}\includegraphics[width=3.7in,height=2.8in]{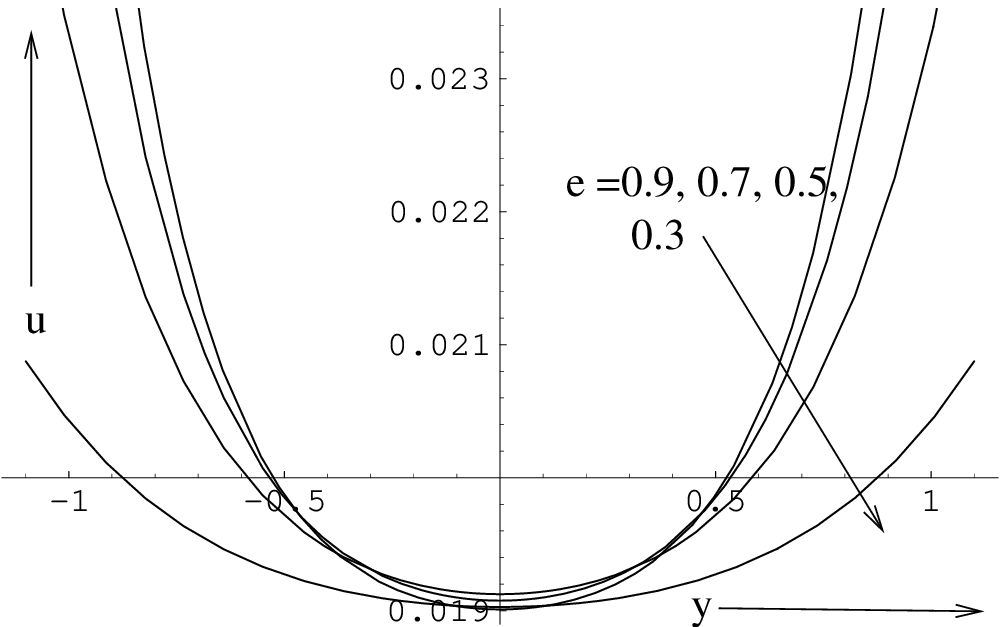}\\$~~~~~~~~~~~~~~~~~~~~~~~~~~~~~~~~~(c)~~~~~~~~~~~~~~~~~~~~~~~~~~~~~~~~~~~~~~~~~~~~~~~~~~~~~~~~~~~~~~~~~~~~~~~~~~~~~~~~~~~~~~~~~~~~~~~(d) ~~~~~~~~~~~~~~$\\ 
Figs. 13: Time averaged mean axial velocity profiles for different values of  e when R=15, $\epsilon$=0.1, $\alpha$=0.25, $\overline{\left(\frac{\partial p}{\partial x}\right)}_2=-2.5$, s=0.0001  (a) k=1; (b) k=0.5; (c)  k=0.1; (d) k=0.05
\end{figure}
\begin{figure}
 \includegraphics[width=3.5in,height=2.6in]{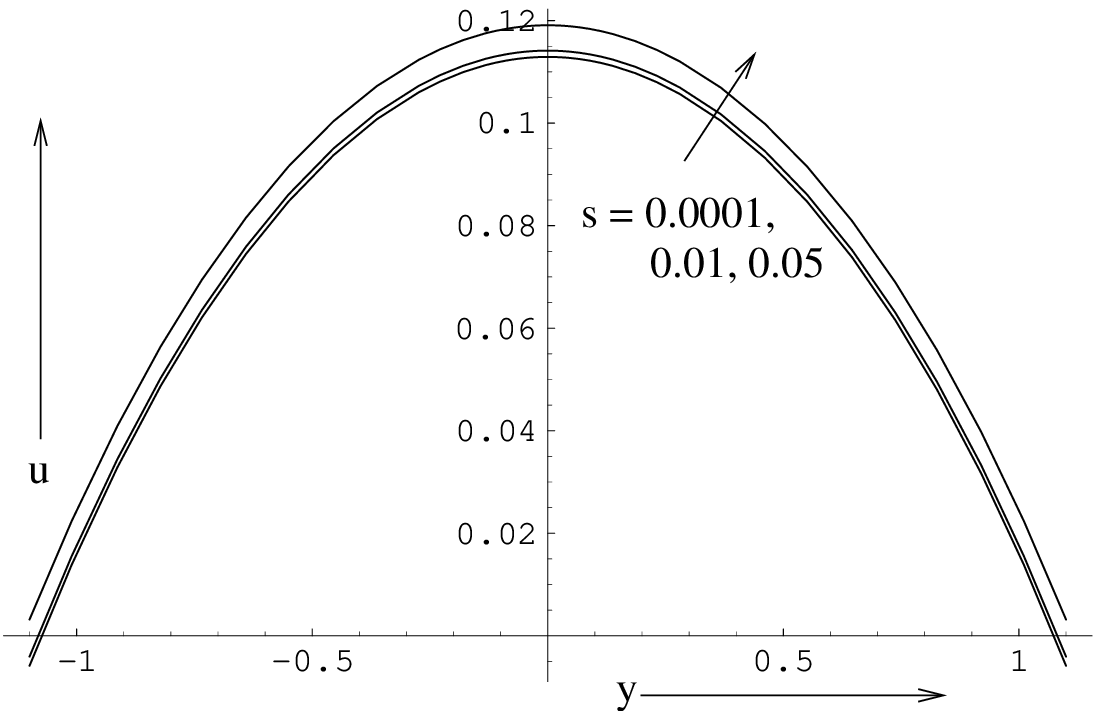}\includegraphics[width=3.5in,height=2.6in]{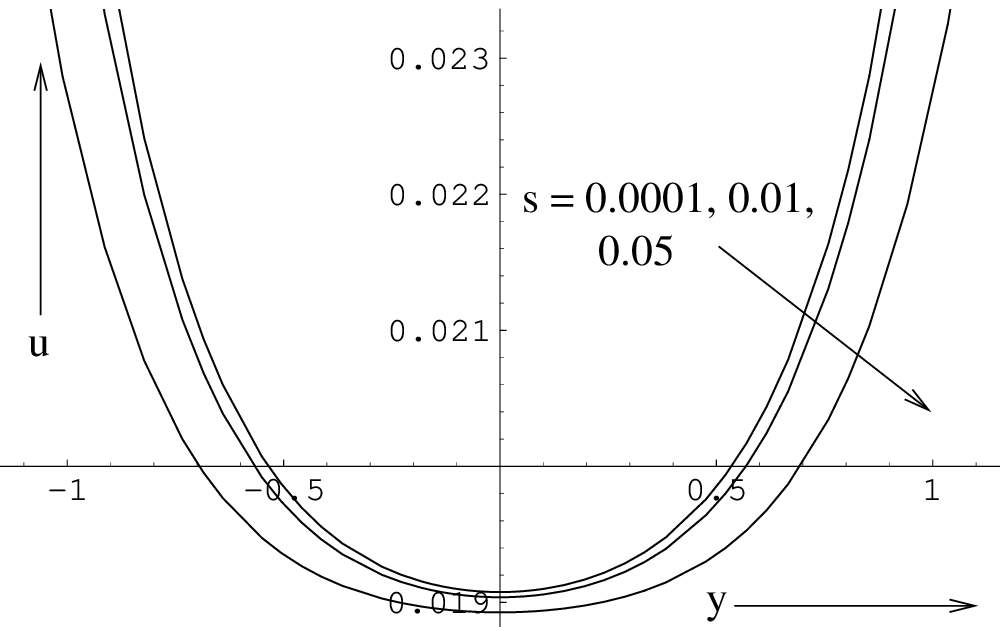}\\$~~~~~~~~~~~~~~~~~~~~~~~~~~~~~~~~~~~~(a)~~~~~~~~~~~~~~~~~~~~~~~~~~~~~~~~~~~~~~~~~~~~~~~~~~~~~~~~~~~~~~~~(b)~~~~~~~~~$\\  
Figs.: 14: Time averaged mean axial velocity profiles for different values of  s when R=15, $\epsilon$=0.1, $\alpha$=0.25, $\overline{\left(\frac{\partial p}{\partial x}\right)}_2=-2.5$, e=0.7  (a) k=1; (b) k=0.05 
\end{figure}
Fig. 13 depicts the variation of the velocity profiles with the
porosity parameter $e$ for fixed values of Reynolds number, pressure
gradient, slip parameter, amplitude ratio and wave number. The
velocity is found to increase, as the porosity parameter increases in
the case when the Darcy number exceeds the value 0.05, though its
parabolic nature changes for small Darcy number. Thus in the situation
as the number of stones decreases, the bile velocity gradually
increases and in the absence of any stone, the bile velocity will be
the greatest. The occurrence of a reverse trend is observed when the
Darcy number equals 0.05. Fig. 14 shows that velocity-slip has a
strong influence on the axial velocity in the porous case. From
Fig. 15, it is observed that wave number affects the axial velocity
more prominently in the vicinity of the boundary.  

\begin{figure}
\begin{center}
 \includegraphics[width=4.0in,height=3.0in]{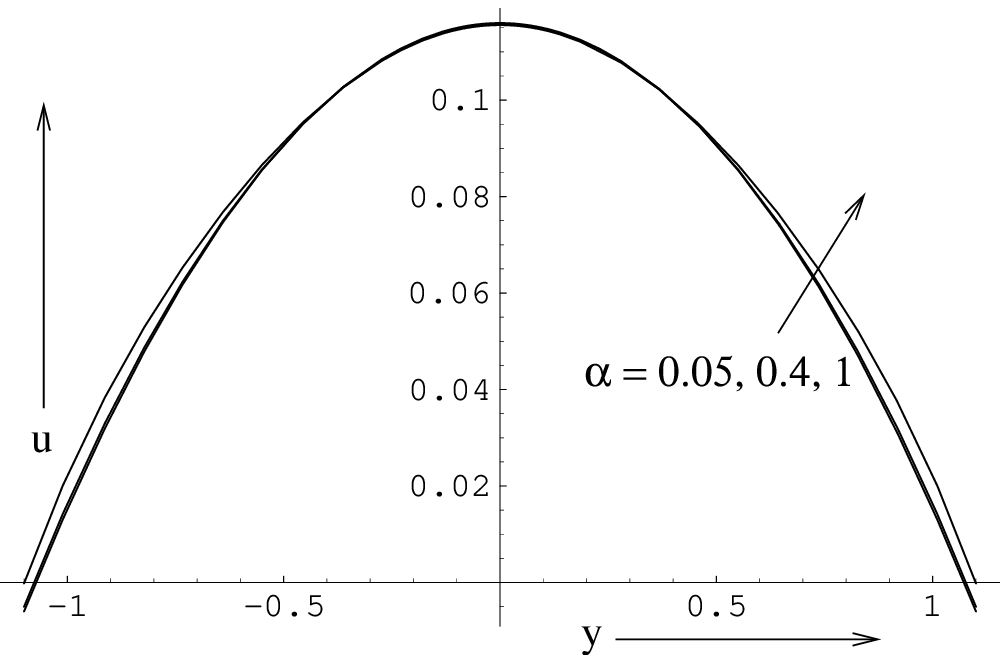}\\
  Figs. 15: Time averaged mean axial velocity profiles for different values of   $\alpha$ when R=15, $\epsilon$=0.1, $\overline{\left(\frac{\partial p}{\partial x}\right)}_2=-2.5$, e=0.7, k=100, s=.001 \\ 
\end{center}
\end{figure}
\begin{figure}
 \includegraphics[width=3.5in,height=2.6in]{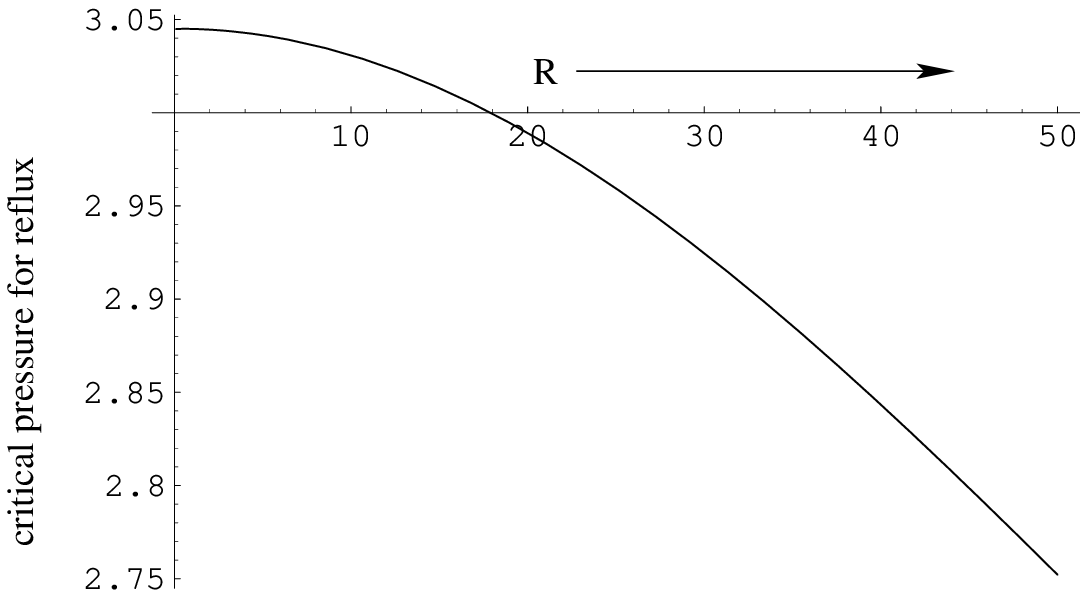}\includegraphics[width=3.5in,height=2.6in]{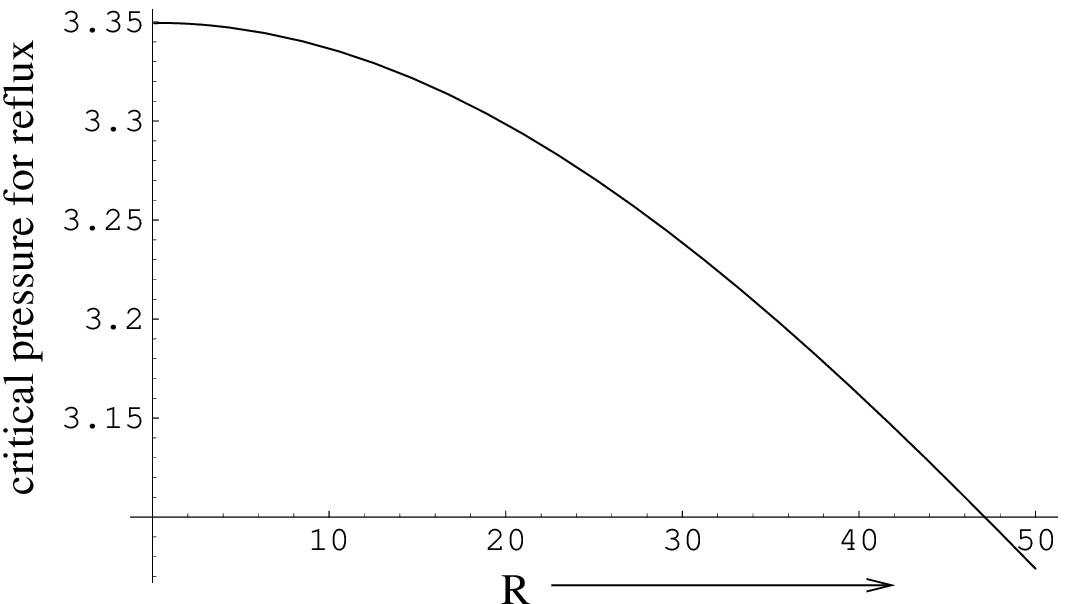}\\$~~~~~~~~~~~~~~~~~~~~~~~~~~~~~~~~~~~(a)~~~~~~~~~~~~~~~~~~~~~~~~~~~~~~~~~~~~~~~~~~~~~~~~~~~~~~~~~~~~~~~~~~~~~~~~~~~~~~~~~~~~~~~~~(b)~~~~~~~~~$\\ 

Figs. 16: Effect of the Reynolds number on  $\overline{\left(\frac{\partial p}{\partial x}\right)}_{2critical~reflux}$ when $\alpha$=0.2, k=1000, s=0.0001 (a) e=0.99; (b) e=0.9
\end{figure}

\begin{figure}
 \includegraphics[width=3.5in,height=2.6in]{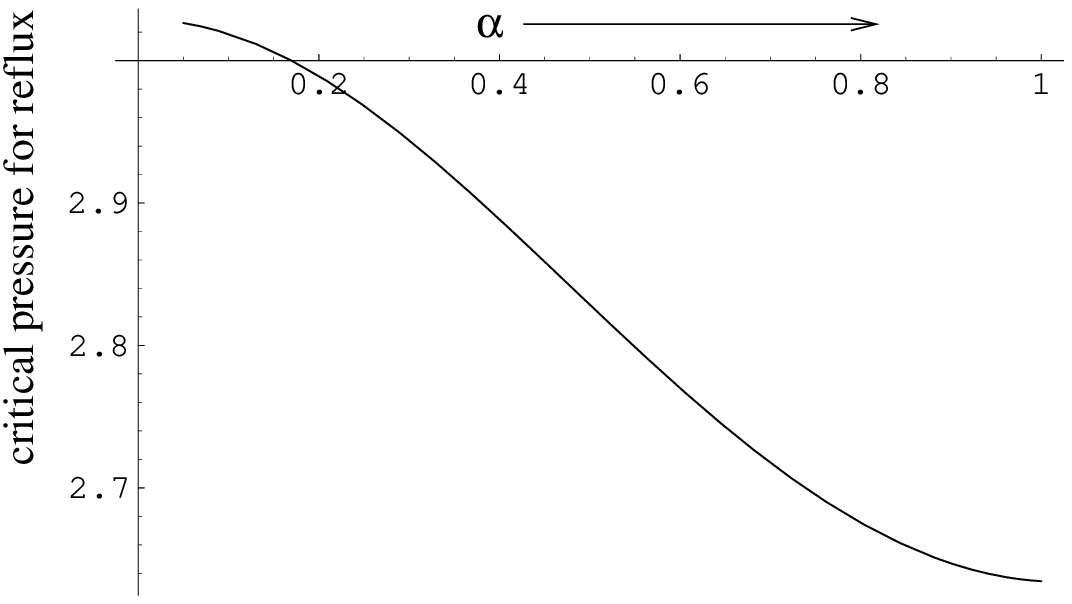}\includegraphics[width=3.5in,height=2.6in]{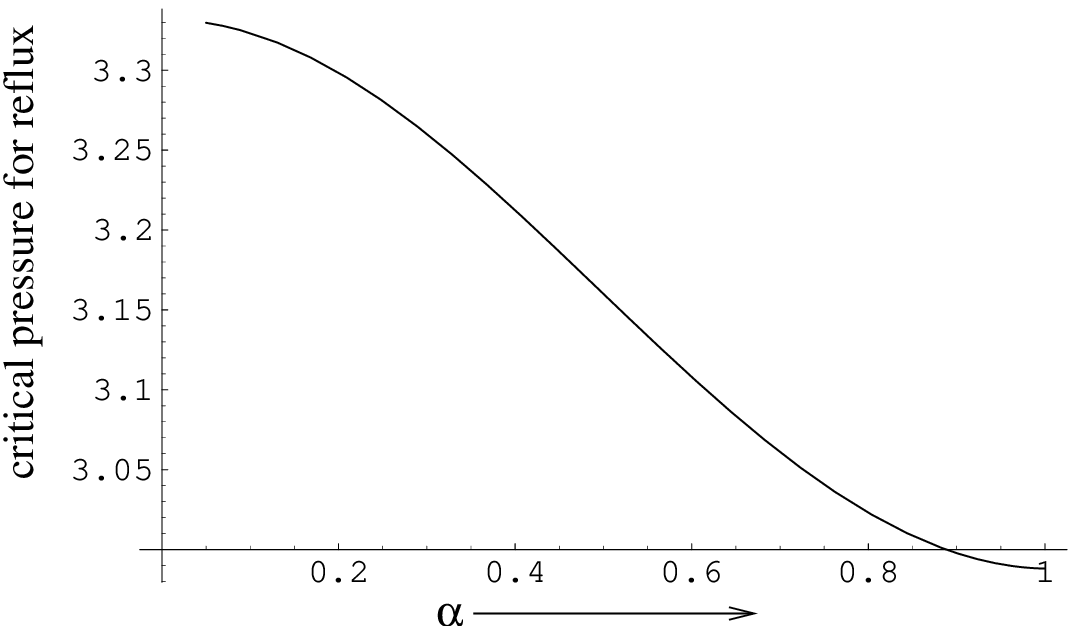}\\$~~~~~~~~~~~~~~~~~~~~~~~~~~~~~~~~~(a)~~~~~~~~~~~~~~~~~~~~~~~~~~~~~~~~~~~~~~~~~~~~~~~~~~~~~~~~~~~~~~~~~~~~~~~~~~~~~~~~~~~~~~(b)~~~~~~~~~$\\ 
Figs. 17: Effect of $\alpha$ on  $\overline{\left(\frac{\partial p}{\partial x}\right)}_{2critical~reflux}$ R=20, k=1000, s=0.0001 (a) e=0.99; (b) e=0.9\\
\end{figure}
\begin{figure}
\includegraphics[width=3.5in,height=2.6in]{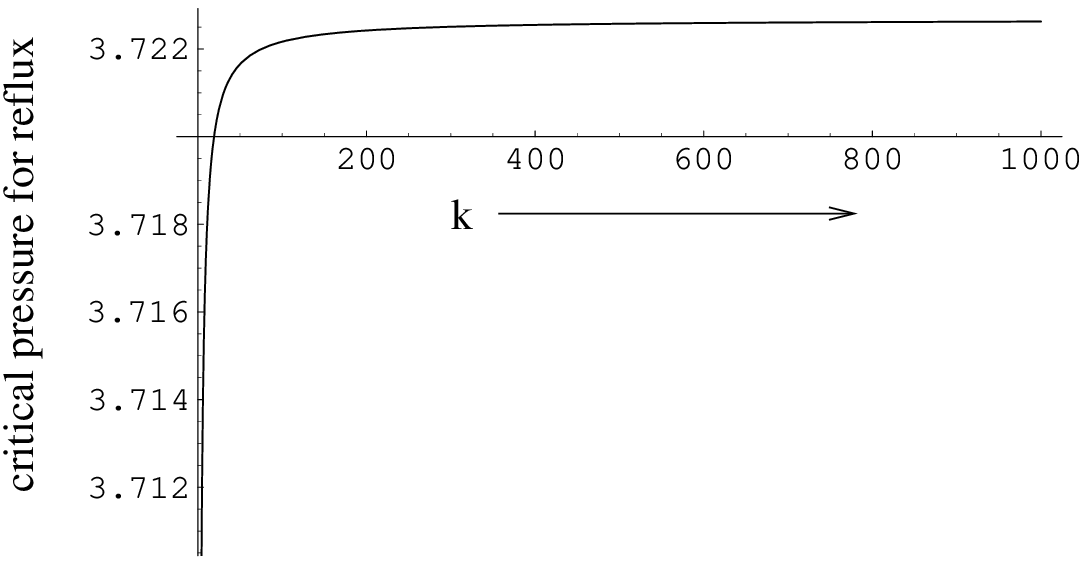}\includegraphics[width=3.5in,height=2.6in]{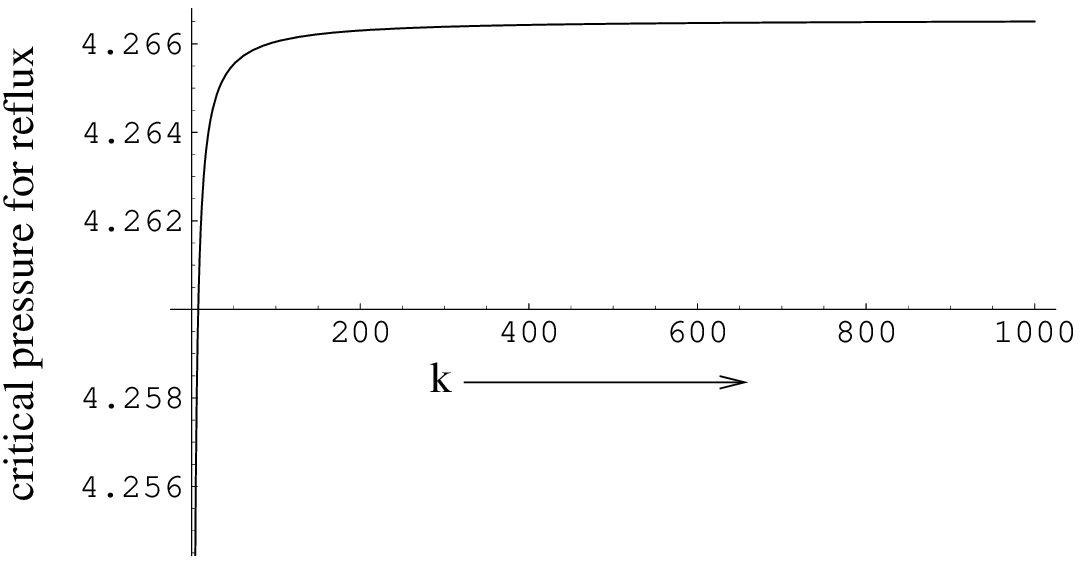}\\$~~~~~~~~~~~~~~~~~~~~~~~~~~~~~~~~~~~~~(a)~~~~~~~~~~~~~~~~~~~~~~~~~~~~~~~~~~~~~~~~~~~~~~~~~~~~~~~~~~~~~~~~~~~~~~~~~~~~~~~~~~~~~~~~~~(b)~~~~~~~~~$\\ 
Figs. 18: Effect of k on  $\overline{\left(\frac{\partial p}{\partial x}\right)}_{2critical~reflux}$ R=20, $\alpha$=0.2, s=0.0001 (a) e=0.8; (b) e=0.7\\
\end{figure}

\subsection{ Critical Pressure for Reflux}

 It is known that bacteria and some other materials sometimes move
 from the bladder to the kidney or from one kidney to the other in the
 direction opposite to the direction of urine flow. This phenomenon is
 referred to as 'ureteral reflux' by physiologists. Severity of
 diseases such as tuberculosis, interstitial cystitis, duct stone
 etc. is often enhanced due to this reflux. It may also happen that
 due to reflux, from the common bile duct, bile flows into the gallbladder
 and is stored there. Results for the critical pressure for reflux
 $R\overline{\left(\frac{\partial p}{\partial
     x}\right)}_{2~critical~reflux}$ computed from equation (40) are
 displayed in Figs. 16-20. Results presented in Figs. 16 match with
 those of Fung and Yih \cite{r2}. This figure illustrates that the
 said critical reflux decreases as the Reynolds number increases. It
 is observed from Figs. 17-19 that the critical pressure reduces with
 an increase in wave number, whereas it first increases with the
 increase in Darcy number and subsequently maintains nearly a constant
 value. It is also worthwhile to observe that the porosity parameter
 significantly affects the magnitude of the critical pressure. From
 Fig. 18, it can be conjectured that when the number of stones in the
 bile is very high, reflux occurs when the critical pressure is
 relatively small. The effect of slip parameter on the critical reflux
 is revealed in Figs. 20. 
\begin{figure}
\begin{center}
 \includegraphics[width=4.0in,height=3.0in]{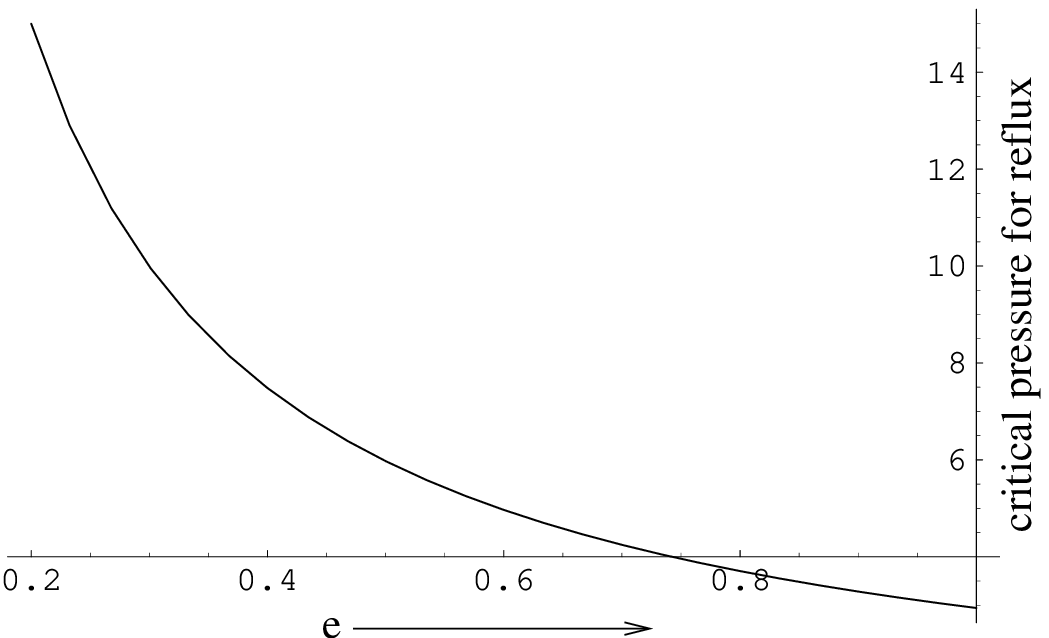}\\
               (Fig. 19) \\ 
Figs. 19: Effect of e on  $\overline{\left(\frac{\partial p}{\partial x}\right)}_{2critical~reflux}$ R=20, $\alpha$=0.2, k=10, s=0.001
\end{center}
\end{figure}
\begin{figure}
 \includegraphics[width=3.5in,height=2.6in]{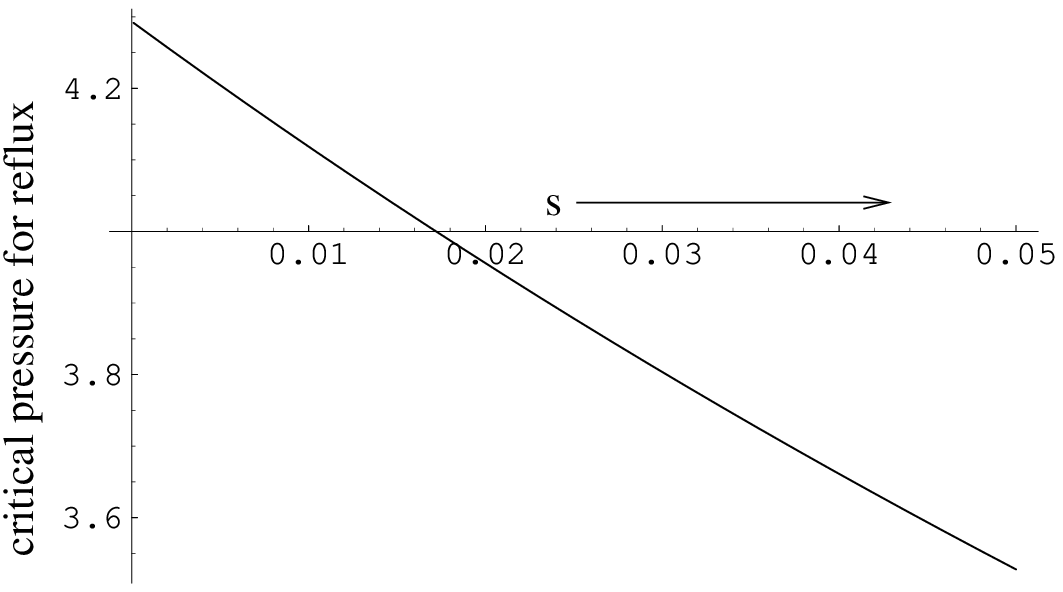}\includegraphics[width=3.5in,height=2.6in]{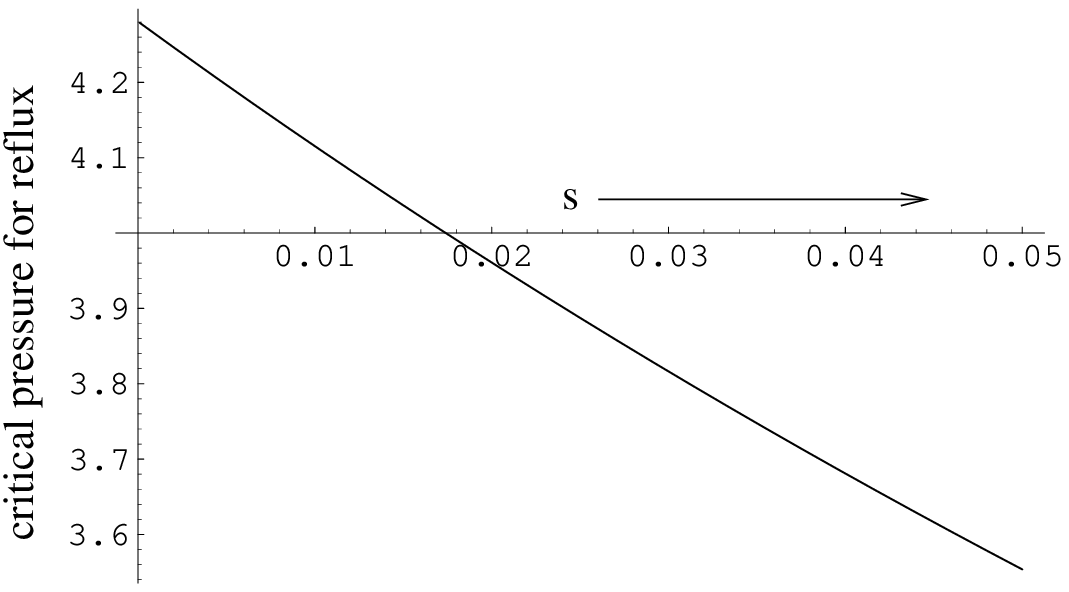}\\$~~~~~~~~~~~~~~~~~~~~~~~~~~~~~~~~~~~~~(a)~~~~~~~~~~~~~~~~~~~~~~~~~~~~~~~~~~~~~~~~~~~~~~~~~~~~~~~~~~~~~~~~~~~~~~~~~~~~~~~~~~~~~~~~~~(b)~~~~~~~~~$\\ 
 \includegraphics[width=3.5in,height=2.6in]{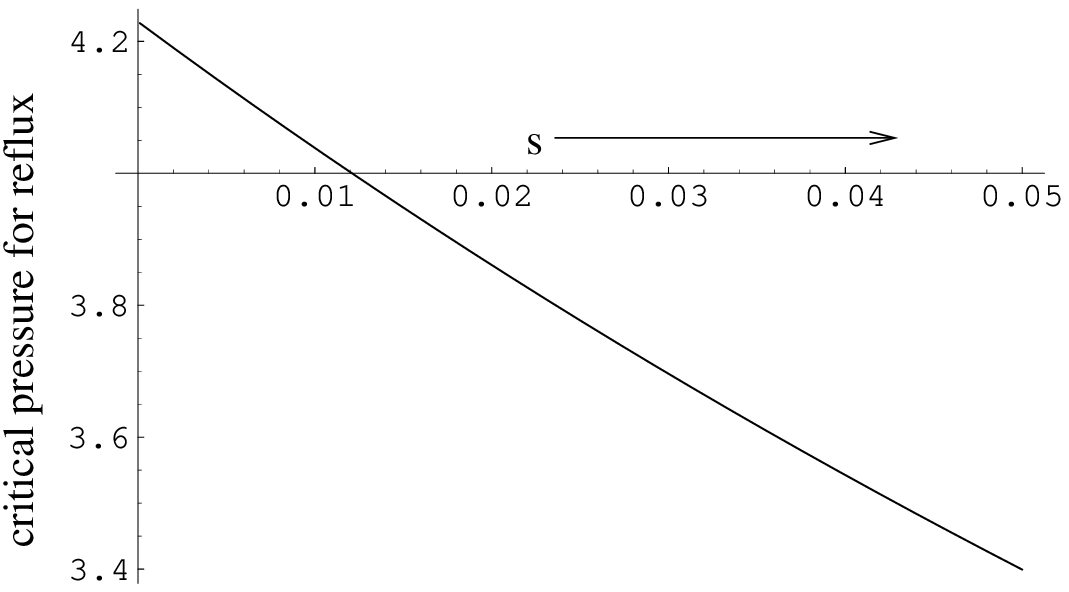}\includegraphics[width=3.5in,height=2.6in]{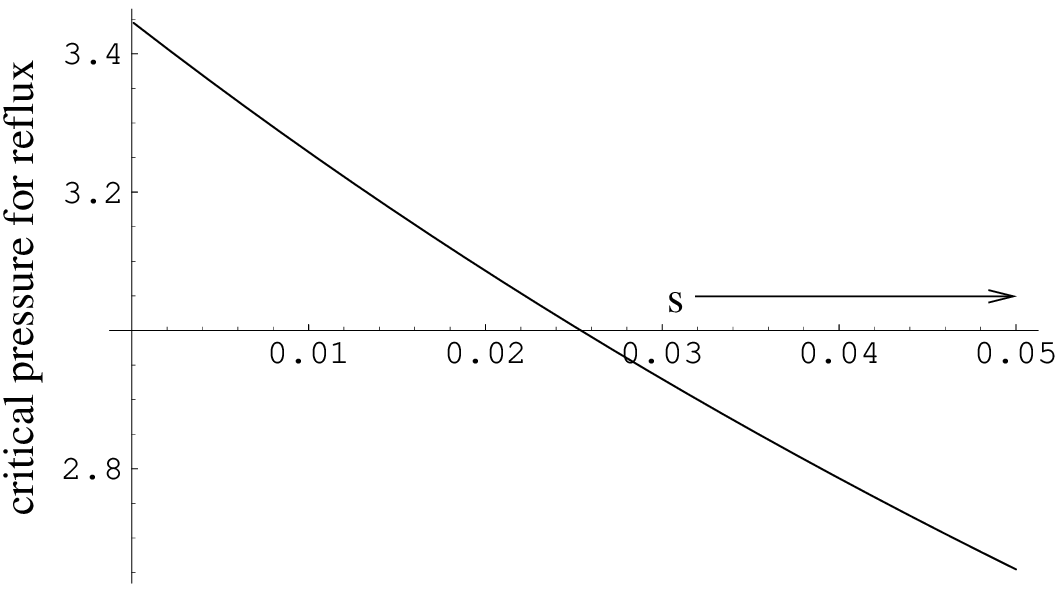}\\$~~~~~~~~~~~~~~~~~~~~~~~~~~~~~~~~~(c)~~~~~~~~~~~~~~~~~~~~~~~~~~~~~~~~~~~~~~~~~~~~~~~~~~~~~~~~~~~~~~~~~~~~~~~~~~~~~~~~~~~~~~~~~~~~~~(d) ~~~~~~~~~~~~~~$\\ 
Figs. 20:  Effect of s on  $\overline{\left(\frac{\partial p}{\partial x}\right)}_{2critical~reflux}$ e=0.7, $\alpha$=0.2, (a) R=10, k=10;  (b) R=15, k=10; (c) R=15, k=1; (d) R=15, k=0.1
\\
\end{figure}

\section{Summary and Conclusion}
The peristaltic transport of a fluid has been investigated here. The
flow is considered to take place in a porous channel. The study
pertains more particularly to a situation where the Reynolds number is
low and the curvature of the channel is quite small. The specific
problem has got relevance to the physiological problem of flow of bile
in the common bile duct
(cf. \cite{r3,r25,r26,r27,r35,r36,r37,r38,r39}). As mentioned in
Section 1, bile flow takes place peristaltically and the common bile
duct is porous. The results presented are applicable to bile flow in
the pathological state, where stones are formed in the bile. Evidence
of slip velocity in bile flow has also been mentioned in Section 1. An
attempt has been made to investigate to some important bio-fluid
dynamical phenomena associated with peristaltic transport.

Quantitative estimates have been presented for the axial velocity and
critical pressure for reflux as a function of Reynolds number,
pressure gradient, porosity parameter, Darcy number,slip parameter,
amplitude ratio as well as wave number. The results and observations
are found to be in good agreement with those reported in
\cite{r2,r33,r34,r35}.\\ From the observations of this study, it
reveals that the velocity profile strongly depends on several
parameters, viz. the mean pressure gradient, porosity, Darcy number as
well as amplitude ratio. The study further indicates that the reflux
will occur if pressure gradient attains a certain critical value.

The present theoretical investigation motivated towards the
peristaltic transport of bile in the bile duct in the presence of
stones can serve as a model that bears the potential to enrich our
understanding of the related physiology of the problem. The important
conjectures that can be made out of the study are as follows : \\(i)
Bile velocity decreases as the stones increases. \\(ii) When bile
contains a very large number of stones, reflux occurs when the
critical pressure is quite small.  \\

{\bf Acknowledgment:} {\it The authors wish to express their deep
  sense of gratitude to all the reviewers for their kind appreciation
  and the estimated comments on the work. One of the authors (Somnath Maiti) is thankful to the Council of Scientific and Industrial Research (CSIR), New Delhi for their financial support towards this study.}


\begin{thebibliography}{99}
\singlespacing
\bibitem {r1} Shapiro, A.H., Jaffrin, M.Y.  and Weinberg,
  S.L., ``Peristaltic pumping with long wavelength at low Reynolds
  number'', J. Fluid Mech. 37 (1969), 799-825.
\bibitem {r2}  Fung Y.C.and Yih C.S., ``Peristaltic Transport'',
  J. Appl. Mech. 35 (1968), 669-75.
\bibitem {r3} Daniel, E. E.; Tomita, T and Watanabe, M., ``Sphincters:
  Normal Function - Changes in Disease'', chapt. 13 (1992), Oddi: Pathophysiology (by J. Toouli and G.T.P.Saccone), page no.201.
\bibitem {r4} Usha, S. and Rao, A. R., ``Effect of curvature and inertia on the peristaltic transport in a two fluid system'', Int. J. Engng. Sci., 38 (2000), 1355-75.
\bibitem {r5} Mishra, M. and Rao, A.R.``Peristaltic transport of a
  Newtonian fluid in an asymmetric channel '', ZAMP. 54 (2003), 532-50.
\bibitem {r6} Mishra, M. and Rao, A.R., ``Peristaltic transport of a power law fluid in a porous tube'', J.Non-Newtonian Fluid Mech. 121
  (2004), 163-74.
\bibitem {r7} Misra, J.C. and Pandey, S.K., ``Peristaltic transport of a
  non-Newtonian fluid with a peripheral layer'',
  Int. J. Engng. Sci. 37 (1999), 1841-58.
\bibitem {r8} Misra, J.C. and Pandey, S.K., ``Peristaltic flow of a
  multi layered power-law fluid through a cylindrical tube'',
  Int. J. Engng. Sci. 39 (2001), 387-402.
\bibitem {r9} Misra, J.C. and Pandey, S.K., ``Peristaltic transport of a
  particle-fluid suspension in a cylindrical tube'',
  Comput. Math. Appl. 28 (1994), 131-45.
\bibitem {r10} Misra, J.C. and Pandey, S.K., ``Peristaltic transport of
  blood in small vessels: study of a mathematical model'',
  Comput. Math. Appl.,43 (2002), 1183-93.
\bibitem {r11} Misra, J.C and Pandey, S.K, ``Peristaltic transport
  in a tapered tube'', Math. Compu. Model., 22(8) (1995), 137-151
\bibitem {r12} Misra, J.C., Maiti S., Shit G.C, ``Peristaltic Transport of a
  Physiological Fluid in an Asymmetric Porous Channel in the Presence
  of an External Magnetic Field'', J. Mech. Med. Biol., 8(4) (2008), 507-525.
\bibitem {r13} Eytan, O., Jaffa, A.J. and Elad, D., ``Peristaltic flow in a tapered channel: application to embryo transport within the
  uterine cavity'', Med. Engng. Phy. 23 (2001), 473-82.
\bibitem {r14} Taylor, G.I., ``Analysis of the swimming of microscopic
  organisms'', Proc. Roy. Soc. Lond. A  209 (1951), 447-61.
\bibitem {r15} Pozrikidis, C.''A study of Peristaltic flow'',  J.Fluid
  Mech. 180 (1987), 515-27.
\bibitem {r16} Carew,E.O and Pedley, T.J., ``An active membrane model for peristaltic pumping: Part I- Periodic activation waves in an infinite tube'', J. Biomech. Eng. 119 (1997), 66-76.
\bibitem {r17} Antanovskii, L.K and Ramkissoon, H., ``Long-wave  peristaltic transport of a compressible viscous fluid in a finite
  pipe subject to a time-dependent pressure drop'', Fluid Dynamics
  Research. 19 (1997), 115-123.
\bibitem {r18} Vries, K.D., Lyons, E.A., Ballard, J., Levi, C.S. and
  Lindsay, D.J., ``Contractions of the inner third of myometrium'',
  Am.J.Obstetries Gynecol. 162 (1990), 679-82.
\bibitem {r19} Keener, J.R. and Sneyd, J., ``Mathematical Physiology'', Springer, 1998.
\bibitem {r20} Berrgel, D.H., ``Cardiovascular Fluid Dynamics'',
  Academic Press, London, 1972. 
\bibitem {r21} Li, A., Nesterov, N.I. Malikova,S.N. and Kiiatkin,
  V.A.``The use of an impulse magnetic field in the combined of
  patients with stone fragments in the upper urinary tract'', Vopr
  kurortol Fizioter Lech Fiz Kult 3 (1994), 22-24.
\bibitem {r22} Torsoli, A. and Ramorino, M. L., ``Motility of the Biliary Tract'', Rendiconti Romani di Gastro Enterologica 2 (1970), 67-80.
\bibitem {r23}	Everhartet J.E, Khare M, Hill M and Maurer
  K.R., ``Prevalence and ethnic differences in gallbladder disease in the United States'', Gastroenterology 117 (1999), 632-9.    
\bibitem {r24}	Khuroo M.S, Mahajan R, Zargar S.A, Javid G and Sapru
  S., ``Prevalence of biliary tract disease in India: A sonographic study in adult population in Kashmir'', Gut 30 (1989), 201-5.  
\bibitem {r25}	Bennion L.J and Grundy S.M., ``Risk factors for
  development of cholelithiasis in man'', N Eng J Med 299 (1978), 1221-7.     
\bibitem {r26}	Maclure K.M, Hayes K.C, Colditz G.A, Stampfer M.J,
  Speizer F.E and Willet W.C., ``Weight, diet and the risk of symptomatic gallstones in middle aged women'', N Engl J Med, 321 (1989), 563-9.    
\bibitem {r27}	Diehl A.K., ``Epidemiology and natural history of gallstone disease'', Gatroenterol Clin North Am, 20 (1991), 1-19.   
\bibitem {r28} Dienstag, JA; Isselbacher, KJ. Tumors of the Liver and Biliary Tract. In: Braunwald E, Fauci AS, Kasper DL, et al., editors. Harrison Principles of Internal Medicine. 15th International Ed. New Delhi: McGraw Hill; 2001. p. 591.
\bibitem {r29} Lauga, E. and Stone, H., ``Effective slip in
  pressure-driven Stokes flow'', J. Fluid Mech. 489 (2003), 55.
\bibitem {r30} Gottschalk, M., Lochner, A., ``Behaviour of postoperative viscosity of bile fluidfrom T-drainage. A contribution to cholelithogenesis'', Gastoenterologisches Journal 50 (2) (1990), 65-67.
\bibitem {r31} Coene, P.P.L.O., Groen, A.K., Davids, P.H.P., Hardeman,
  M.Tytgat, G.N.J., Huibregtse, K., `` Bile viscosity in patients with
  biliary drainage: Effect of co-trimoxazole and N-acetylcysteine and
  role in stent clogging'', Scandinavian Journal of Gastroenterology
  29 (1994), 757-763.
\bibitem {r32} Luo, X.Y., Chin, S.B., Ooi, R.C., Clubb, M., Johnson,
  A.G., Bird, N., The rheological properties of human bile, Fourth
  International Conference on Fluid Mechanics. Dalian, China (2004).
\bibitem {r33} Toouli J., ``Sphincter of Oddi'', Gastroenterologist 4
  (1996); 44-53.
\bibitem {r34} Torsoli A., ``Physiology of the human sphincter of Oddi'', Endoscopy 20(1) (1988), 166-170.
\bibitem {r35} Sugita R., Sugimura E., Itoh M., Ohisa T., Takahashi
  S. and Fujita N., ``Pseudolesion of the bile duct caused by flow effect: a diagnostic pitfall of MR cholangiopancreatography'',
  Am. J. Roentgenol 180 (2003), 467-471.
\bibitem {r36} Toouli J., Dodds W.J., Honda R., Sarna S., Hogan W.J.,
Komarowski R.A. and Linehan J.H., ``Arndorfer RC. Motor function of the opossum sphincter of Oddi'', J. Clin. Invest. 71 (1983), 208-220.
\bibitem {r37} Calabuig R., Ulrich-Baker M.G, Moody F.G. and Weems
  W.A., ``The propulsive behavior of the opossum sphincter of Oddi'', Am. J. Physiol. 258 (1990), G138-G142.
\bibitem {r38} Grivell M.B, Woods C.M, Grivell A.R, Neild T.O, Craig A.G,
Toouli J. and Saccone G.T., ``The possum sphincter of Oddi pumps or
resists flow depending on common bile duct pressure: a multilumen manometry study'', J. Physiol., 558 (2004), 611-622.
\bibitem {r39} Toouli J., Geenen J.E, Hogan W.J, Dodds W.J and
  Arndorfer R.C., ``Sphincter of Oddi motor activity: a comparison between patients with common bile duct stones and
controls'', Gastroenterology 82 (1982), 111-117.
\end{thebibliography}
\end{document}